\newif\ifapjtemplate
\apjtemplatetrue

\ifapjtemplate
  \documentclass[11pt,onecolumn]{aastex701}
\else
  \documentclass{article}
  \usepackage[left=1in, right=1in, marginparwidth=1in]{geometry}
  \usepackage{authblk}
\fi

\usepackage{amsmath, amsthm, amsfonts, amssymb}
\usepackage{graphicx}
\usepackage{xcolor}
\usepackage{mathtools}
\usepackage{comment}
\usepackage{hyperref}

\ifapjtemplate
  \makeatletter
  \def\email@error#1{}
  \def\correspondingauthor#1{{\ifanonymous
   \else\global\@correspauthortrue\global\correspauthortrue
  \gdef\@correspondingauthor{#1}
  \fi}
  }
  \def\@email[#1]#2{\ifanonymous
  \else\ifcorrespauthor
  {\onecolumngrid\renewcommand\thefootnote{}\addtocounter{footnote}{-1}
  \footnote{\hskip-11ptCorresponding author: \@correspondingauthor, \href{mailto:#2}{#2}\ifmodern\vrule depth 7pt width 0pt\relax\else\vrule depth 7pt width0pt\fi}}
  \else
  \expandafter\gdef\csname email@\romannumeral\the\allauthorcnt\endcsname{}
  \ifthenelse{\equal{#1}{show}}{
  \protected@xdef\@@email{\ifx\@@email\empty#2\else\@@email, #2\fi}}{}
  \fi\fi\global\correspauthorfalse}
  \makeatother
\else
  \usepackage{natbib}
\fi

\newcommand{\reals}{\mathbb{R}}

\ifapjtemplate
  \renewcommand{\tensor}[1]{#1}
\else
  \newcommand{\tensor}[1]{#1}
\fi
\newcommand{\spacedomain}{\mathcal{X}}
\newcommand{\facedomain}{\mathcal{Y}}
\newcommand{\angledomain}{\Omega}

\newcommand{\ghost}{\mathfrak{g}}
\newcommand{\emissi}{\varepsilon}

\newcommand{\orcidauthor}[2]{
  \if\relax\detokenize{#1}\relax
    \author{#2}
  \else
    \author[#1]{#2}
  \fi
}

\newcommand{\AlexOrcid}{0000-0003-3152-8206}
\newcommand{\PatrickOrcid}{0000-0003-2131-4634}
\newcommand{\AdityaOrcid}{0009-0009-8285-861X}
\newcommand{\JoshuaOrcid}{0000-0003-4353-8751}
\newcommand{\ChadOrcid}{0000-0002-7530-6173}
\newcommand{\JonahOrcid}{0000-0001-6432-7860}
\newcommand{\LukeOrcid}{0000-0001-7364-7946}

\newcommand{\setupfrontmatter}{
  \title{Tensor Networks for Block Adaptive Thermal Radiation Transport}
  \ifapjtemplate
    \shorttitle{Tensor Networks for Block Adaptive TRT}
    \shortauthors{Gorodetsky et al.}
    \orcidauthor{\AlexOrcid}{Alex A. Gorodetsky$^\dagger$}
    \affiliation{Department of Aerospace Engineering, University of Michigan, Ann Arbor, MI}
    \affiliation{Michigan SPARC, Los Alamos National Laboratory, Ann Arbor, MI}
    \correspondingauthor{Alex A. Gorodetsky}
    \email{goroda@umich.edu}
    {\renewcommand\thefootnote{}\addtocounter{footnote}{-1}\footnote{\hskip-11pt$^\dagger$\,Both co-lead authors provided equal contributions.}}
    \orcidauthor{\PatrickOrcid}{Patrick D. Mullen$^\dagger$}
    \affiliation{Michigan SPARC, Los Alamos National Laboratory, Ann Arbor, MI}
    \affiliation{Computing and Artificial Intelligence, Los Alamos National Laboratory, Los Alamos, NM}
    \orcidauthor{\AdityaOrcid}{Aditya S. Deshpande}
    \affiliation{Department of Aerospace Engineering, University of Michigan, Ann Arbor, MI}
    \affiliation{Michigan SPARC, Los Alamos National Laboratory, Ann Arbor, MI}
    \orcidauthor{\JoshuaOrcid}{Joshua C. Dolence}
    \affiliation{Michigan SPARC, Los Alamos National Laboratory, Ann Arbor, MI}
    \affiliation{Computing and Artificial Intelligence, Los Alamos National Laboratory, Los Alamos, NM}
    \orcidauthor{\ChadOrcid}{Chad D. Meyer}
    \affiliation{Michigan SPARC, Los Alamos National Laboratory, Ann Arbor, MI}
    \affiliation{Continuum Models and Numerical Methods, Los Alamos National Laboratory, Los Alamos, NM}
    \orcidauthor{\JonahOrcid}{Jonah M. Miller}
    \affiliation{Michigan SPARC, Los Alamos National Laboratory, Ann Arbor, MI}
    \affiliation{Computing and Artificial Intelligence, Los Alamos National Laboratory, Los Alamos, NM}
    \orcidauthor{\LukeOrcid}{Luke F. Roberts}
    \affiliation{Michigan SPARC, Los Alamos National Laboratory, Ann Arbor, MI}
    \affiliation{Computing and Artificial Intelligence, Los Alamos National Laboratory, Los Alamos, NM}
  \else
    \author[1,2]{Alex A. Gorodetsky\thanks{goroda@umich.edu}}
    \author[2,3]{Patrick D. Mullen}
    \author[1,2]{Aditya S. Deshpande}
    \author[2,3]{Joshua C. Dolence}
    \author[2,4]{Chad D. Meyer}
    \author[2,3]{Jonah M. Miller}
    \author[2,3]{Luke F. Roberts}
    \affil[1]{\normalsize Department of Aerospace Engineering, University of Michigan, Ann Arbor, MI}
    \affil[2]{\normalsize Michigan SPARC, Los Alamos National Laboratory, Ann Arbor, MI}
    \affil[3]{\normalsize Computing and Artificial Intelligence, Los Alamos National Laboratory, Los Alamos, NM}
    \affil[4]{\normalsize Continuum Models and Numerical Methods, Los Alamos National Laboratory, Los Alamos, NM}
    \date{\vspace{-10pt}}
  \fi
}

\begin{document}
\setupfrontmatter

\ifapjtemplate\else\maketitle\fi

\begin{abstract}
Thermal radiation transport (TRT) is computationally challenging because the photon specific intensity depends simultaneously on space, angle, and frequency, producing a high-dimensional state space whose cost grows rapidly with resolution. Tensor-network methods, such as the tensor train (TT) decomposition, can mitigate this curse of dimensionality when the solution admits low-rank structure, but localized nonseparable structure can drive substantial rank growth. Adaptive mesh refinement (AMR) provides a complementary means of making high spatial resolution tractable by concentrating resolution on features of interest that would otherwise be prohibitively expensive to resolve. When tensor representations are constructed locally on AMR blocks, this decomposition offers an additional advantage: nonseparable structure can remain confined to the blocks in which it occurs rather than driving rank growth throughout the domain. This paper introduces a framework for combining block-adaptive mesh refinement with block-local tensor-train representations of the photon specific intensity. The approach couples spatial adaptivity with localized low-rank compression, allowing fine-scale features to be resolved while limiting the impact of locally complex structure on tensor ranks elsewhere in the domain. We develop conservative ghost-cell transfer, coarse-fine flux correction, and refinement criteria that balance spatial discretization and tensor-rounding errors. Numerical experiments demonstrate that the method preserves low-rank structure across much of the domain while resolving localized features with AMR, achieving compression factors exceeding $\sim$10$^3$. The resulting framework provides a practical foundation for applying tensor-network methods to high-dimensional TRT on the multiscale, adaptively refined meshes encountered in radiation hydrodynamics simulations.
\end{abstract}

\section{Introduction}\label{sec:intro}

Thermal radiation transport plays a central role in a wide range of astrophysical and high-energy-density systems. Important examples include star formation~\citep{wunsch2024radiation}, black hole accretion~\citep{zhang1, zhang3, zhang2}, line-driven outflows~\citep{proga2000,proga2004,dyda23}, and inertial confinement fusion~\citep{haines23,SAUPPE2024101156,ko24}. In many such settings, the radiation field both influences and responds to spatially complex and dynamically evolving flow structures, making accurate transport a central component of predictive multiphysics simulation.

Several numerical strategies are used to solve radiation transport equations, including discrete ordinates ($S_N$) methods, Monte Carlo methods, and moment-based closures such as flux-limited diffusion and M1~\citep{imc71,levermore81,m1,davis2012,jiang21,jiang22}. In this work we focus on $S_N$ methods, which evolve the angle-resolved specific intensity directly and therefore avoid both the sampling noise inherent to Monte Carlo methods and the closure assumptions associated with reduced moment systems~\citep{brunner02,foucart15}.
This direct angular resolution makes $S_N$ attractive in regimes where anisotropy, shadowing, and transport across mixed optical-depth conditions are important, but it comes at substantial computational cost because the radiation field must be resolved over space and angle, and in multigroup settings over frequency as well~\citep{jiang21,jiang22}.

In many applications, however, the challenge is not only high dimensionality in angle and frequency, but also the need to resolve localized spatial structure within a larger multiphysics calculation. Radiation transport is often coupled to (magneto)hydrodynamics in settings with sharp material interfaces, shocks, and evolving multiscale features. For such problems, block-structured adaptive mesh refinement is a central numerical tool for concentrating resolution where it is needed while keeping the global calculation tractable, and modern multiphysics codes are increasingly organized around such frameworks~\citep[e.g.,][]{paramesh,pluto,amrex,grete23,athenapp,stone24}.

At the same time, thermal radiation transport often remains one of the most expensive components of these simulations because of its stiffness and its dependence on space, angle, and, in multigroup settings, frequency~\citep{southworth24,jiang21,jiang22,tttt,mgtttt}. For transport solvers based on $S_N$, this creates an additional numerical requirement beyond angular and spectral fidelity: the method must remain compatible with block-structured refinement, inter-block data exchange, and conservative treatment across coarse-fine interfaces.

In previous work, we developed global tensor-network step-then-truncate solvers for gray and multigroup thermal radiation transport~\citep{tttt,mgtttt}. These studies showed that substantial reductions in storage and computational cost can be obtained when the specific intensity exhibits exploitable multilinear structure. Similar opportunities have been identified in other high-dimensional PDE settings, including the Fokker--Planck equation~\citep{dolgov2012fast}, Vlasov equations~\citep{kormann2015semi,einkemmer24}, and quantum dynamics~\citep{soley2021functional}. Nevertheless, practical deployment of such methods requires compatibility with the block-structured AMR frameworks used in modern multiphysics simulations. 

In this paper, we develop a block-local tensor-train representation of the specific intensity for thermal radiation transport on block-structured adaptive meshes. We describe how information is transferred between blocks through ghost-cell communication and coarse-fine flux correction, and we formulate refinement strategies that expose the interaction between tensor rounding and spatial refinement. We then apply the method to several model problems to assess its compression behavior, conservation properties, and practical performance.

Related work has explored adaptive tensor representations in several forms. Localized refinement of tensor-product approximations has been considered for multivariate function approximation, where refinement is confined to local tensor patches rather than propagated through a single global tensor product~\citep{peterka23adaptive}. Adaptive patching strategies have also been proposed for tensor-train computations to exploit localized structure through divide-and-conquer decomposition~\citep{grosso26patching}. In addition, adaptive stochastic Galerkin finite element methods have been combined with global hierarchical tensor representations to couple physical-mesh refinement with adaptive stochastic resolution~\citep{eigel17adaptive}. However, these approaches do not address the setting considered here: a conservative block-structured $S_N$ transport solver in which tensor representations must support ghost-cell communication, coarse-fine flux correction, and mesh adaptation while remaining aligned with the underlying AMR decomposition.

The rest of the paper is organized as follows. In \S\ref{sec:background}, we introduce the governing equations and summarize the numerical structure needed for the block-structured AMR setting. In \S\ref{sec:tensors}, we review the tensor-train representation and the tensor operations needed for blockwise communication. In \S\ref{sec:tt_mesh_refinement}, we present the block-local tensor-train AMR formulation, including ghost-cell transfer, flux correction, and refinement criteria. In \S\ref{sec:results}, we present numerical examples.  In \S\ref{sec:discussion} we provide a brief discussion where we analyze the interaction between rounding and (a) spatial refinement and (b) conservation.  We provide a brief conclusion in \S\ref{sec:conclusion}.

\section{Background}\label{sec:background}

\subsection{Governing equations}

Let $\spacedomain  \subset \reals^d$ denote a spatial domain, $\Omega = [0, 2\pi] \times [0 , \pi]$ denote the angular domain, and let $\nu \in \reals_+$ denote frequency. In this paper we assume a static background material with mass density $\rho$; local thermal equilibrium (LTE); and isotropic, elastic scattering. Furthermore, we consider the multigroup approach where the frequency domain is divided into $N_\nu$ bins with end points $\nu_0,\nu_1,\ldots,\nu_{N_{\nu}}$ satisfying $\nu_0=0$ and $\nu_{N_{\nu}}=\infty.$ Under these assumptions, the group intensity $I_f(t,x,\theta,\phi) : \reals_+ \times \spacedomain \times \Omega \to \reals_+$ associated with the frequency bin $[\nu_{f-1}, \nu_f)$ satisfies~\citep{mihalas1984}
\begin{equation}
\partial_t I_f + c \mathbf{n} \cdot \nabla I_f = c\left[ \rho \kappa_{s, f} \left(J_f - I_f\right) + \rho \kappa_{a, f} (\emissi_f - I_f) \right]
\label{eq:multigroup}
\end{equation}
where $c$ is the speed of light; $\mathbf{n}$ is a unit normal coordinate direction; $\kappa_{s, f}$ is the specific group mean scattering opacity; $\kappa_{a, f}$ is the specific group mean absorption opacity; $J_f = \frac{1}{4\pi} \int I_f d\Omega$ is the group mean intensity; and $\emissi_f$ is a group emission coefficient related to the Planck function $B(\nu, T)$ according to
\begin{equation}
\emissi_{f} = \frac{c}{4\pi} \int_{\nu_{f-1}}^{\nu_f} B(\nu, T) d \nu \quad \text{ and } B(\nu, T) = \frac{8\pi h \nu^3}{c^3} \frac{1}{\exp(h \nu / [kT])-1},
\end{equation}
where $h$ is  the Planck constant. The specific case corresponding to a single group $[\nu_0, \nu_{N_\nu}) = \reals_+$ corresponds to the gray approximation, where $\emissi=\emissi_\mathrm{gray} = ca T^4 / [4\pi]$ and $a$ is the radiation constant. 
Finally, the thermal radiation transport problem is coupled to the background fluid through a material temperature equation
\begin{equation}
\partial_t T = - \sum_{f} \left[\frac{ c\kappa_{a, f} }{c_v} \left( \frac{4 \pi}{c} \emissi_f - E_f \right)\right],
\end{equation}
where $c_v$ is the specific heat capacity at constant volume of the material and $E_f = \frac{4 \pi}{c} J_f$ is the group radiation energy density.

\subsection{Discretization and Numerics} 
In this section, we summarize the discretization and numerical structure needed for the AMR framework. Our emphasis is on the boundary communication and conservation issues that arise once the transport solve is distributed over blocks, rather than on the detailed source-term or transport discretizations. Although the numerical examples use the solver developed in \cite{tttt} and \cite{mgtttt}, the domain-decomposition and AMR ideas described below apply more broadly to tensor-network transport solvers. The gray setting on a 2D spatial domain is sufficient to expose the main issues while keeping the notation manageable, and the key elements carry over directly to the multigroup setting in 3D.

Within this context, consider a two-dimensional spatial global domain $\spacedomain \subset \reals^2.$ Suppose this spatial domain is discretized into $N_x \times N_y$ uniform cells with centers $\hat{\spacedomain} \subset\mathcal{X}$. These cells are {\it interior} to the domain, and later we will add exterior ghost cells. Furthermore, assume that the angular space is discretized into $N_\theta \times N_\phi$ angles with values $\hat{\Omega} \subset\Omega.$ Let us denote a spatial multi-index $i \in \mathcal{I} \equiv [0,\ldots, |\hat{\spacedomain}|-1]$ to identify a unique spatial cell. Similarly $\ell \in[0,\ldots, N_\theta-1]$ and $p \in [0,\ldots,N_\phi-1]$ denote unique angular indices. $\hat{I}^m \in \reals^{N_x \times N_y \times N_\theta \times N_\phi}$ denotes the approximation of the cell- and angle- averaged intensity on this grid at some timestep $m$.

Next, let $\hat{\facedomain}^{x/y}$ denote the $x/y$-faces of the cells, and $F^x \in \reals^{(N_x + 1) \times N_y \times N_\theta \times N_\phi}$ denote a discretized field defined on the $x$-faces. To enforce boundary conditions, we use a standard ghost-cell based approach. First we extend our domain in all directions to $\hat{\spacedomain}_{ext}$ and assume that all ghost cells $\hat{\spacedomain}_\ghost := \hat{\spacedomain}_{ext} \setminus \hat{\spacedomain}$ are specified by a problem boundary condition operator at each timestep $m$, $\mathcal{B}^m: \hat{\spacedomain} \times \hat \Omega \rightarrow \hat{\spacedomain}_\ghost \times \hat \Omega$, satisfying $\mathcal{B}^m_{i \ell p}[\cdot] = 0 \, \forall \, i \in \mathcal{I}$. Then, at the beginning of each timestep, the ghost cells are filled in by applying this boundary operator to the interior solution $\hat{I}^m$ from the previous time-step:
\begin{equation}
\hat I^{m}_{ext} = \texttt{GhostPad}[\hat I^{m}] + \mathcal{B}^{m+1}[\hat I^{m}],
\end{equation}
where $\texttt{GhostPad}$ is a function that takes $\hat{I}^m$ defined on $\hat{\spacedomain}$ and appends zeros in the ghost cells. We write this process as a sum because that is the form naturally followed by a tensor-network implementation~\citep{tttt}. When $\hat{I}^m$ is represented in a structured format, one does not manipulate entries directly. Instead, elementwise updates must be recast as algebraic operations, as discussed in \S\ref{sec:tensors}. 

Once boundary conditions are fixed, various solvers can be deployed to obtain the solution at the next time step.  At a high-level we solve the discretized system for the solution at the next time-step in the domain interior using a splitting-based scheme for the following,
\begin{equation}
\hat I^{m+1} - \hat I^{m} +\mathcal{T}[\hat I^m_{ext}] = \mathcal{S}[\hat I^{m+1}],
\end{equation} 
where $\mathcal{T}: \hat{\spacedomain}_{ext} \times \hat \angledomain \rightarrow \hat{\spacedomain} \times \hat \angledomain$ is the non-spatially local, explicit transport operator and $\mathcal{S}: \hat{\spacedomain} \times \hat \angledomain \rightarrow \hat{\spacedomain} \times \hat \angledomain$ is the spatially-local, implicit emission, absorption, and scattering operator. Since $\mathcal{S}$ is spatially local, decomposing the spatial domain has no impact on how we treat it relative to our previous work and we discuss it no further. We can further decompose the transport operator in terms of flux operators.  Fluxes are calculated on faces from the intensity via the operator $\mathcal{F}^{x/y}: \hat \spacedomain_{ext} \rightarrow \hat \facedomain^{x/y}$ \citep[see][for details]{tttt}. Next, consider face-to-cell shift operators $\mathcal{M}^{x\pm}: \hat \facedomain^x \rightarrow \hat \spacedomain$, defined component-wise as $\mathcal{M}^{x\pm}_{i, j \ell p}[F^x]= F^x_{i\pm 1/2, j \ell p}$. The purpose of face-to-cell shift operators is to select values of a face field on the right/left side of a cell. Using these objects, the transport operator can be applied in two steps. First, the flux on the faces is calculated
\begin{equation*}
\hat F^x = \mathcal{F}^x[\hat I^m_{ext}].
\end{equation*}
Then the flux on each face is manipulated to determine the total amount of quantity coming in and out of the cell through the boundaries
\begin{equation*}
\mathcal{T}^x[\hat I^m_{ext}] = \frac{\Delta t}{\Delta V} \left(\mathcal{M}^{x+}[\Delta A^x\hat F^x] - \mathcal{M}^{x-}[\Delta A^x\hat F^x]\right), 
\end{equation*}
where $\Delta A^x \in \reals^{(N_x + 1) \times N_y}$ is a field of face areas on $\hat \facedomain^x$ and $\Delta V \in \reals^{N_x \times N_y}$ is field holding the volumes of the cells. Clearly, this choice maintains exact conservation of the volume integral of $\hat I$ (up to changes induced by $\mathcal{S}$). 

\subsection{Domain Decomposition}\label{sec:mesh_refinement}

\begin{figure}[t]
\centering
\includegraphics[width=0.7\textwidth]{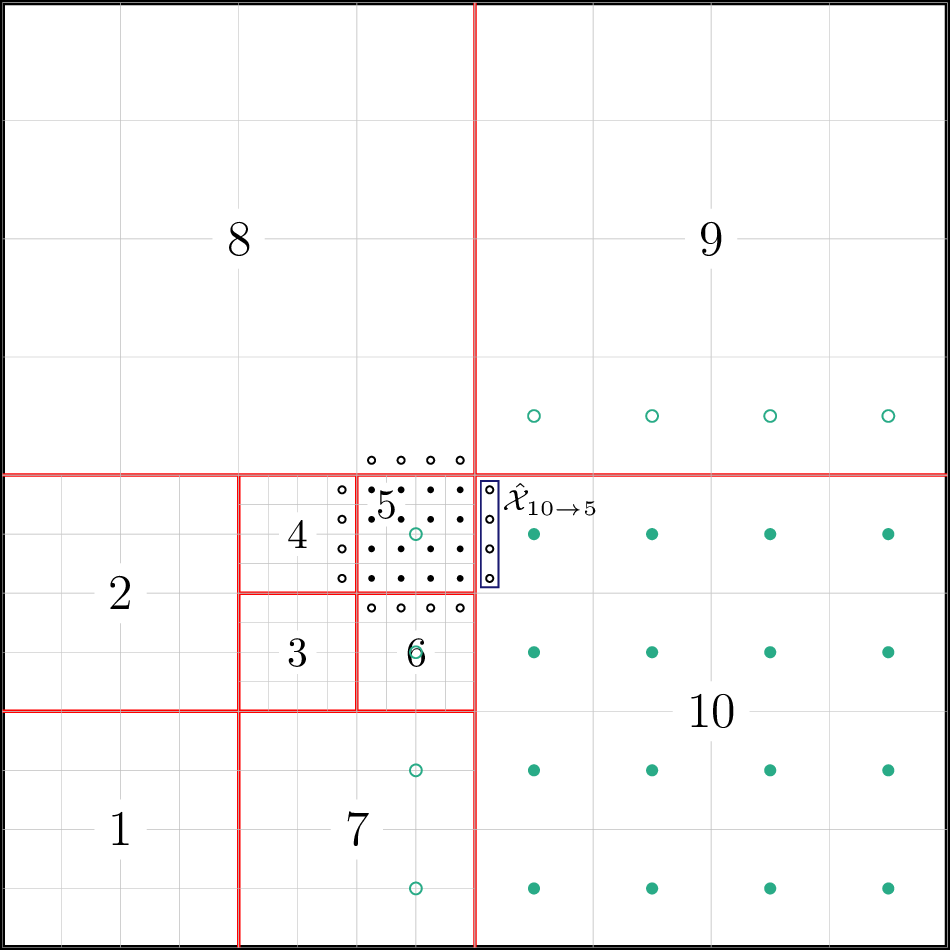}
\caption{Exemplar domain decomposition scenario. Each domain is labeled and has an equal number of cells. The interior cell centers for domain 5 and 10 are shown in solid circles of black and green, respectively. The ghost cells, those that are interior to the domain, are shown in open black and green circles. Corner ghost cells are not shown since they are not used in the computation. The blue  rectangle shows ghost nodes of domain $5$ that are informed by domain $10$, $\hat{\spacedomain}_{10\to5}.$}
\label{fig:block_mesh}
\end{figure}

Our AMR construction is built around a blockwise spatial decomposition that preserves the full angular discretization within each block. Instead of a single uniform discretization $\hat{\spacedomain}$, we partition the domain into $N_b$ contiguous, non-overlapping blocks of cells $\{\hat \spacedomain_b\}$, such that $\cup_{b} \hat \spacedomain_b = \hat{\spacedomain}$. An example is shown in Figure~\ref{fig:block_mesh}, where the domain is partitioned into $N=10$ labeled blocks of equal logical size. In this work, we consider only non-overlapping blocks, i.e., $\hat{\spacedomain}_b \cap \hat{\spacedomain}_c = \emptyset$ for $b\neq c$. Let $\hat{I}_b \in \reals^{N_{x} \times N_\theta \times N_\phi}$ denote the approximation of the cell-averaged intensity over block $b$, where $I_b:\spacedomain_b \times \angledomain \to \reals$. Throughout the discussion below, we assume that each block contains the same number of spatial cells and that all blocks share the same angular discretization. In principle, one could allow different spatial and angular resolutions across blocks, but differing angular resolutions would require additional interpolation during communication. Here, only the spatial domain is decomposed, and each block retains the full angular discretization.

Blocks communicate with their neighbors through ghost-cell filling and, when refinement levels differ, through a conservative flux correction on shared interfaces. Once this communication is completed for a given time step, the solver can proceed independently within each block. The process closely mirrors the single-block case, with the additional requirement of coarse-fine flux reconciliation.

Each block is assigned a halo of ghost cells, $\hat \spacedomain_{b,\ghost}$, analogous to the undecomposed problem. Some of these ghost cells lie on the physical boundary $\hat \spacedomain_{ext} \setminus \hat \spacedomain$, while others overlap the interiors of neighboring blocks $\hat \spacedomain_c$. In Figure~\ref{fig:block_mesh}, for example, the interior cell centers for domains 5 and 10 are shown as filled circles, and their ghost cells within the global domain are shown as open circles. We also define blockwise cell faces $\hat \facedomain_{b}^{x/y}$ analogously to the undecomposed problem, together with blockwise operators for fluxes, source terms, and shifts, as well as blockwise cell-volume and face-area fields. With this notation, one time step on the decomposed mesh consists of three conceptual stages: filling ghost data, computing and correcting fluxes, and updating cell-averaged intensities.

\paragraph{Fill ghost data} To fill a block's ghost cells at each time step, we first define a block-local boundary operator $\mathcal{B}_b^{m}: \{\hat \spacedomain_c \times \hat \angledomain\}_{c \in \text{neighbors}(b)} \rightarrow \hat \spacedomain_{b, \ghost} \times \angledomain$,
which copies or prolongates cell values from neighbor blocks on interior ghost regions and sets the boundary conditions from $\mathcal{B}^m$ on physical boundary ghost regions. The interior ghost cells are filled through boundary operators for each neighbor block $c$, $\mathcal{B}^m_{c \rightarrow b}: \hat \spacedomain_c \times \hat{\angledomain} \rightarrow \hat \spacedomain_{b,\ghost} \times \hat{\angledomain}$. 
 In the interior of $b$ we just have $\mathcal{B}^m_{c \rightarrow b, i \ell p}[\cdot] = 0, \forall \, i \in \hat{\spacedomain}_{b}$. The operations performed in the overlap region between the ghosts of $b$ and the interior of $c$, $\hat{\spacedomain}_{b,\ghost} \cap \hat{\spacedomain}_c$, depend on the discretization levels of the neighbors. When the overlapping cells on the blocks are in one-to-one correspondence, $\hat{\spacedomain}_{b,\ghost} \subset \hat{\spacedomain}_c$ (note the hat on block $c$), then $\mathcal{B}_{c \rightarrow b}$ just copies the values from $c$ to $b$. When a ghost cell comprises multiple cells on the neighbor block, the values of the finer cells are averaged, i.e., restriction is performed. When a ghost cell is finer than the neighbor block cells, its value is conservatively prolongated from the coarser neighbor cells.
To enable the preservation of exact tensor operations and to ensure stability, we will only consider prolongation approaches that can be represented by linear spatial operators.  A linear operator representation of prolongation is denoted by $\mathcal{R}_{c\to b}:\hat{\spacedomain}_c \to \hat{\spacedomain}_{b,\ghost} \cap \hat{\spacedomain}_c.$ Importantly, the same prolongation procedure is performed for every angle, which means that the overall solution prolongation is multilinear.
Our numerical examples will use the simplest possible piecewise constant prolongation operator for $\mathcal{B}_{c \rightarrow b}$ at coarse-fine boundaries.

Using these neighbor boundary operators, we fill in each block's ghost cells again through a summation over the neighboring physical boundaries and interior neighbors. 
\begin{equation}
\hat I^{m}_{b,ext} = \texttt{GhostPad}[\hat I_b^{m}] + \mathcal{B}^{m+1}_{b, \text{physical}}[\hat I^{m}_b] + \sum_{c \in \text{neighbors}(b)} \mathcal{B}_{c \rightarrow b}^{m+1}[\hat I_c^{m}].
\label{eq:boundary_tt}
\end{equation}
 
\paragraph{Flux computation and correction}
After ghost cells for each block are filled, the flux is evaluated as
\begin{equation*}
\hat F^x_b \leftarrow \mathcal{F}^x_b[\hat I^m_{b,ext}].
\end{equation*}
 Note that neighbor blocks have $\hat \spacedomain_b \cap \hat \spacedomain_c = \emptyset$ but $\hat \facedomain_{b}^{x} \cap \hat
  \facedomain_{c}^{x} \neq \emptyset$. Therefore, any field defined on $\{\hat{\facedomain}_{1}^{x},...\hat{\facedomain}_{N}^{x}\}$,
  such as the flux, admits multiple representations on block boundaries. Our scheme is conservative only if these overlapping
  representations are made consistent. In particular, at a boundary between a coarse and a fine block there are multiple fine faces
  corresponding to a single coarse face.
  Since we take the face fluxes to be area-averaged quantities, conservation must be enforced on the face-integrated flux $\Delta A^x \hat F^x$.
  Although it is possible to construct $\mathcal{B}_{b \rightarrow c}$ and $\mathcal{F}_b$ so that this correspondence is satisfied by
  construction, in practice it is more convenient to overwrite the shared coarse-face flux with the area-weighted restriction of the
  fine-face fluxes.
  We therefore define the flux correction operator on shared $x$-faces, $\mathcal{C}^x_{c \rightarrow b}:
  \{\hat \facedomain^x_c \times \angledomain, \hat \facedomain^x_b \times \angledomain\} \rightarrow \hat \facedomain^x_b \times
  \angledomain$ by
  \begin{equation}
  \mathcal{C}^x_{c \rightarrow b,i \ell p}[\hat F^x_c, \hat F^x_b]
  =
  \mathcal{R}[\Delta A_c^x \hat F^x_c]_{b,i \ell p}
  -
  \Delta A^x_{b,i}\hat F^x_{b,i \ell p},
  \qquad
  i \in \hat \facedomain_{b}^{x} \cap \hat \facedomain_{c}^{x},
  \end{equation}
  where $\mathcal{R}$ denotes area-weighted restriction from the fine grid of $c$ to the coarse grid of $b$. The coarse-face integrated
  flux is then corrected according to
  \begin{equation}
  \Delta A_b^x \hat F^x_b
  \leftarrow
  \Delta A_b^x \hat F^x_b
  +
  \sum_{c \in \text{finer neighbors}(b)} \mathcal{C}^x_{c \rightarrow b}[\hat F^x_c, \hat F^x_b].
  \label{eq:flux_correct}
  \end{equation}
An analogous correction is applied on shared $y$-faces.
\paragraph{Update cell-averaged intensities}
After flux correction, the chosen solver can proceed independently within each block. Again, this is represented by solving
\begin{equation*} 
\hat I_b^{m+1} - \mathcal{S}_b[\hat I^{m+1}] =  \hat I_b^m - \frac{\Delta t}{\Delta V_b} \left[
\left(\mathcal{M}^{x+}_b[\Delta A^x_b\hat F^x_b] - \mathcal{M}^{x-}_b[\Delta A^x_b\hat F^x_b] \right)
+
\left(\mathcal{M}^{y+}_b[\Delta A^y_b\hat F^y_b] - \mathcal{M}^{y-}_b[\Delta A^y_b\hat F^y_b] \right)
\right].
\end{equation*}

\subsection{Tensor-train decomposition}\label{sec:tensors}

Our method represents the specific intensity in each block using a tensor-train (TT) ansatz~\citep{oseledets2011tensor}. The main purpose of this section is to identify the TT operations that make blockwise AMR possible, particularly those needed for ghost-cell communication, flux correction, and local refinement. For additional details on gray and multigroup transport solvers in TT form, we refer to our prior work~\citep{tttt,mgtttt}. Other tensor-network representations are also possible, but the AMR mechanisms developed here are tailored to the TT structure.

Tensor networks are motivated by the fact that a standard cell-centered representation of the (gray) photon specific intensity in 2D requires storing $N_xN_y N_\theta N_\phi$ variables. For the rest of this paper we will refer to the number of discretized cells per block by $N_x$, regardless of the dimensionality. Furthermore, we split these variables into three groups: space, $\theta$, and $\phi$. After such a splitting, the specific intensity is represented as a product of three {\it TT-cores}:
\begin{equation}
 \hat{I}_b[i, \ell, p] = X_{b}[i, :] \Theta_b[:, \ell, :] \Phi_b[:, p],
\label{eq:tensor_decomp}
\end{equation}
where $X_b \in \reals^{N_x  \times r_b^1}$ represents the spatial variables\footnote{For multigroup problems with $N_\nu > 1$, this would become a ``spatiospectral" core \citep[see][]{mgtttt}}, $\Theta_b \in \reals^{r^1_b \times N_\theta \times r^2_b}$, and $\Phi_b \in \reals^{r^2_b \times N_\phi}$, and \texttt{MATLAB} notation is used to index certain rows and columns. The two outside cores are matrices, and the middle core is a 3-way tensor. The sizes of these cores $r^i_b$ are called {\it ranks}.  We will sometimes write
\begin{equation}
    \hat{I}_b = X_b \otimes \Theta_b \otimes \Phi_b
\end{equation}
to represent the full tensor generated by the elementwise operation~\eqref{eq:tensor_decomp}.

When represented in this format, the storage requirement of each block drops from $N_x N_\theta N_\phi$ to $r^1_b N_x + r^1_b r^2_b N_\theta + r^2_b N_\phi$.

The most important tensor-train operation that is required for ghost cell communication is that of addition.  Consider two three-dimensional tensors $\tensor{T}$ and $\tensor{S},$ with corresponding cores $(T_1,T_2,T_3)$ and $(S_1,S_2,S_3)$, respectively. Addition of tensors in TT format yields a tensor $V$ with the following format
\begin{align}
    V[i, \ell, p] &= T[i, \ell, p] + S[i, \ell, p] = 
    \underbrace{\begin{bmatrix}
    T_{1}[i, :] & S_{1}[i, :]
    \end{bmatrix}}_{V_{1}[i, :]}
    \underbrace{
    \begin{bmatrix}
        T_{2}[:,\ell, :] & 0 \\
        0 & S_{2}[:,\ell,:]
    \end{bmatrix}
    }_{V_{2}[:,\ell,]}
    \underbrace{
    \begin{bmatrix}
    T_{3}[:, p] \\
    S_{3}[:, p]
    \end{bmatrix}
    }_{V_{3}[:, p]}
    = V_{1}[i, :]V_{2}[:,\ell,:] V_{3}[:, p],
\end{align}
with the ranks of $V$ being the {\it sum} of the corresponding ranks of $T$ and $S$ \citep{tttt}. Moreover, we see that this additive structure yields a block-diagonal structure for the central core. 

In the step-and-truncate solver considered here, this rank growth must be controlled through a rounding procedure such as TT-SVD~\citep{oseledets2011tensor}. Rounding re-approximates a TT representation by one with lower ranks while incurring a prescribed relative accuracy penalty $\epsilon$. In practice, this operation often becomes the dominant cost because its computational complexity scales like $\mathcal{O}(dnr^3)$. 

To place these operations in context, suppose that the solution in a given block and its neighbors has ranks bounded by $r$. In two dimensions, a block has four neighbor sets, so filling its ghost cells requires the addition of five tensors in total, which yields a rank-$5r$ representation. Computing the flux then applies a multilinear operator of rank $2$ \citep[see][]{tttt}, producing a rank-$10r$ tensor. Naive rounding of that tensor therefore has cost $\mathcal{O}(1000dnr^3)$. More specialized strategies, such as Gram-SVD~\citep{al2022parallel} and randomized rounding~\citep{al2023randomized}, can exploit the block-diagonal structure of sums to reduce this cost. Custom Gram-SVD variants can additionally exploit sparsity in the spatial fibers.

\section{Block-local tensor-train mesh refinement}\label{sec:tt_mesh_refinement}

This section describes the central contribution of the paper: a block-local AMR procedure that preserves the TT representation while maintaining conservative block communication.

\subsection{Block updates}
We now describe how to implement~\eqref{eq:boundary_tt} and the flux correction~\eqref{eq:flux_correct}. The first step is to define $\texttt{GhostPad}$. Adding ghost cells and padding them with zeros corresponds to adding zero rows to the spatial core,
\begin{equation}
    \hat{I}_{b,\text{pad}} =  \texttt{GhostPad}(\hat{I}_b) = 
    \underbrace{
    \begin{bmatrix}
        \textbf{0}_{\texttt{nghost} \times r_b^1} \\ 
        X_b        
    \end{bmatrix}
    }_{X_{b,\text{pad}}}
    \otimes \Theta_b \otimes \Phi_b
\end{equation}
where $\texttt{nghost}$ is the number of ghost cells for the $b$-th block. The corresponding rows of the core then map to the ghost cell indices. After padding we know that 
\begin{equation}
    \hat{I}_{b,\text{pad}}[i,j,k] = 0 \quad \text{ for } \quad i=0,\ldots,\texttt{nghost}-1,\  j = 0,\ldots, N_\theta-1,\ \text{ and } \ k=0,\ldots,N_\phi -1.
\end{equation}
The next step is to add the contributions from the neighbors. Consider a single neighbor, either an interior block or the physical boundary. The update has the form
\begin{equation}
    \hat{I}_{b,\text{up}} = \hat{I}_{b, \text{pad}}  + T_{c\to b},
\end{equation}
where $\tensor{T}_{c\to b} \in \reals^{\texttt{nghost} \times N_\theta \times  N_\phi}$ transfers information from domain $c$ to domain $b$. Let $\mathcal{I}_{c\to b}$ denote the subset of ghost-cell indices in $\hat{I}_{b,\text{up}}$ that are informed by domain $c$. Then this tensor update satisfies
\begin{equation}
    \tensor{T}_{c\to b}[i,j,k] = \left\{ \begin{array}{cl}
       t_{c\to b}[i,j,k] & \textrm{ if } i \in \mathcal{I}_{c\to b}, \  j=0,\ldots,N_\theta-1,\ \text{ and } \ k=0,\ldots,N_\phi -1 \\
       0 & \text{ otherwise},
  \end{array}
    \right.
\end{equation}
where $t_{c\to b} \in \reals^{\texttt{nghost} \times N_\theta\times  N_\phi}$ is a tensor representing the new ghost-cell values. This tensor is nonzero only on $\mathcal{I}_{c\to b} \subset [\texttt{nghost}]$, since domain $c$ contributes only to a subset of block $b$'s ghost cells. All remaining entries of $\tensor{T}_{c \to b}$ are zero. The tensor $t_{c \to b}$ is derived from $\hat{I}_c$ and represented in low-rank form. If the TT representation of this transfer tensor is $t_{c \to b} = X_{c\to b} \otimes \Theta_{c\to b} \otimes \Phi_{c \to b}$, then
\begin{equation}
    \hat{I}_{b,\text{up}} = 
    \begin{bmatrix}
        \textbf{0}_{\texttt{nghost} \times r^1_b} & X_{c\to b} \\ 
        X_b        & \textbf{0}_{N_{x} \times r^1_{c\ \to b} }
    \end{bmatrix}     
    \otimes 
    \begin{bmatrix}
    \Theta_b  & \textbf{0}_{r^1_b \times N_\theta \times r_b^2} \\ 
    \textbf{0}_{r^1_{c\to b} \times N_\theta \times r^2_b} & \Theta_{c \to b}
    \end{bmatrix}
    \otimes 
    \begin{bmatrix}
    \Phi_b  \\
    \Phi_{c\to b}
    \end{bmatrix}.
\end{equation}
This addition requires a block column in the spatial core, a block-diagonal middle core, and a block row in the final angular core. In practice, block $b$ has several neighbors, and each contributes an analogous term. It remains to be determined how $t_{c \to b}$ is constructed.

In this work, the transfer tensors $t_{c \to b}$ are obtained through piecewise constant prolongation and restriction. Nonlinear transfers are in principle possible, but they would require approximate procedures such as cross approximation or problem-specific nonlinear tensor operations. We therefore restrict attention to the linear operators described in \S\ref{sec:mesh_refinement}. Their main advantage is that they act on the spatial and angular cores independently. More specifically, let $\mathcal{R}_{c\to b, \texttt{pad}} \in \reals^{\texttt{nghost} \times N_{x}}$ denote a row-padded version of the transfer operator described in Section~\ref{sec:mesh_refinement}. This padding inserts zeros for ghost cells not contained in $\hat{\spacedomain}_{b,\ghost}$. Then $t_{c\to b}$ is obtained by applying $\mathcal{R}_{c\to b,\text{pad}}$ to every column of the first core $X_{c}$ while reusing the angular cores from $\hat{I}_c$:
\begin{equation}
t_{c\to b} = \left(\mathcal{R}_{c\to b, \text{pad}}X_{c}\right) \otimes \Theta_c \otimes \Phi_c.
\end{equation}

The procedure just described can be readily adapted to flux correction, since~\eqref{eq:flux_correct} also consists of sums of tensors. However, performing flux correction in this way would lead to a growth in the number of tensors that need to be summed. In turn, this would cause additional computational expense as described in the end of \S\ref{sec:tensors}. Instead, when we transfer information from the fine blocks to the coarse ones, we perform addition on the {\it flux tensors} $\hat{F}^x$ and $\hat{F}^y$ rather than for the ghost cells. This allows us to use an appropriately scaled fine flux for the faces of the coarse cell. 

\subsection{Adaptivity}
Adaptive refinement changes only the spatial discretization local to a block. Splitting a block therefore requires prolongation of its spatial core onto finer child grids, while merging requires restriction back to a coarser parent grid. In the linear setting considered here, these operations are precisely the transfer operations described above and therefore act only on the spatial core while leaving the angular cores unchanged. This is a key practical advantage of the block-local TT representation: mesh adaptation can be performed without rebuilding the full angular structure of the tensor.

\subsection{Rounding}
\label{sec:rounding}
Given a prescribed global error budget, different allocations of the allowable rounding error lead to different blockwise rounding strategies. In the simplest case, a uniform strategy applies the same relative rounding tolerance to every block. Alternatively, the allowable absolute error may be distributed approximately equally across blocks, which permits larger relative tolerances in regions where the intensity is small. We refer to this as intensity-weighted partitioning. For meshes containing blocks of different physical sizes, the allocation can also account for block volume, giving a volume-intensity partitioning strategy. The corresponding blockwise tolerances and their relationship to the finite-volume discretization error are derived in Appendix \ref{app:error_analysis}.

A limiting version of this idea is useful when large portions of the domain contain negligible radiation. Blocks whose intensity falls below a prescribed threshold may be classified as sparse and rounded very aggressively, effectively reducing them to rank one. This suppresses tensor-rank growth associated with numerically insignificant structure in regions that contribute little to the overall radiation field. Because this approach intentionally relaxes local error control, the threshold should be chosen relative to physically meaningful intensity scales. The practical consequences of these strategies are examined in Section \ref{sec:discussion}.

\subsection{Compression}
Let $\mathcal{S}_{\rm \mathrm{global\text{-}TT}}$ denote the TT storage required to represent the radiation field with a single tensor train spanning the full spatial domain, and let $\mathcal{S}_{\rm \mathrm{block\text{-}TT}}$ denote the total TT storage obtained by representing the solution independently on each block. We define the diagnostic
\begin{equation}
  \mathcal{C}_d
  \equiv
  {\mathcal{S}_{\rm \mathrm{global\text{-}TT}} \over \mathcal{S}_{\rm \mathrm{block\text{-}TT}}}
  \label{eq:cd_definition}
\end{equation}
for quantifying the storage benefit of a block-local tensor representation relative to a single global TT representation.  With this convention, $\mathcal{C}_d>1$ indicates that the block-decomposed representation is more compact than the corresponding global TT representation, while $\mathcal{C}_d<1$ indicates that block decomposition increases the total tensor storage.  This quantity isolates the effect of domain-local tensor compression: it does not compare against the full uncompressed $S_N$ representation, but rather asks whether decomposing the spatial domain exposes additional local low-rank structure relative to a single global tensor train.

For a global tensor train with ranks $r_1$ and $r_2$, the storage is
\begin{equation}
  \mathcal{S}_{\rm \mathrm{global\text{-}TT}}
  =
  r_1 \mathcal{N}_{\rm cells}
  +
  r_1 r_2 N_\theta
  +
  r_2 N_\phi ,
  \label{eq:global_tt_storage_general}
\end{equation}
where $\mathcal{N}_{\rm cells}$ is the \emph{total} number of spatial cells represented by the calculation.  For a block-decomposed representation, the corresponding storage is
\begin{equation}
  \mathcal{S}_{\rm \mathrm{block\text{-}TT}}
  =
  \sum_{b=1}^{N_b}
  \left(
    r_{1,b} \mathcal{N}_{{\rm cells},b}
    +
    r_{1,b} r_{2,b} N_\theta
    +
    r_{2,b} N_\phi
  \right) ,
  \label{eq:block_tt_storage_general}
\end{equation}
where $\mathcal{N}_{{\rm cells},b}$ is the number of spatial cells on block $b$, and $(r_{1,b},r_{2,b})$ are the corresponding local TT ranks.

Equation~(\ref{eq:block_tt_storage_general}) also clarifies what block adaptivity must accomplish in order to be beneficial.  The spatial contribution $\sum_b r_{1,b} \mathcal{N}_{{\rm cells},b}$ can be reduced when local ranks $r_{1,b}$ are smaller than the global rank $r_1$.  However, the angular storage terms $\sum_b (r_{1,b} r_{2,b} N_\theta + r_{2,b} N_\phi)$ sum over blocks, whereas in the global representation Equation~(\ref{eq:global_tt_storage_general}) they appear only once as $(r_1 r_2 N_\theta + r_2 N_\phi)$.  Thus, block decomposition is advantageous only when the reduction in $\sum_b r_{1,b} \mathcal{N}_{{\rm cells},b}$ relative to $r_1 \mathcal{N}_{\rm cells}$ more than offsets the increase from $(r_1 r_2 N_\theta + r_2 N_\phi)$ to $\sum_b (r_{1,b} r_{2,b} N_\theta + r_{2,b} N_\phi)$.  In this sense, quantities such as    
\begin{equation}
  {r_1 \over \langle r_1 \rangle}
  \qquad {\rm and} \qquad
  {r_1 r_2 \over \langle r_1 r_2 \rangle},
  \label{eq:rank_reduction_diagnostics}
\end{equation}
where $\langle \cdot \rangle$ denotes the average over blocks, $\langle r_1 \rangle = N_b^{-1} \sum_b r_{1,b}$, provide useful diagnostics for whether the global TT rank is being driven by localized structure.  Large values of these ratios indicate that the global representation is carrying rank generated by only a subset of the domain, which is precisely the scenario in which a block-local representation can reduce storage.  Conversely, if the local ranks remain comparable to the global ranks everywhere, then the block-decomposed representation offers little rank reduction while incurring additional overhead for each block.  

\section{Results}~\label{sec:results}
We assess the block-adaptive TT framework using several test problems spanning different physical regimes and mesh-refinement strategies. In the following, the rank-1 asymptotic analysis and 2D line-source problems are gray (i.e., frequency integrated); the crooked pipe test is gray with absorption/emission spanning optically thick and thin regimes with sharp material interfaces, and the stellar irradiation problem is multigroup with frequency-dependent absorption.  We consider three mesh-refinement strategies combined with tensor-train decomposition: TT-DD (tensor train with domain decomposition on a uniform grid), TT-SMR (tensor train with static mesh refinement, where the refinement pattern is fixed in time), and TT-AMR (tensor train with adaptive mesh refinement, where blocks are dynamically refined and/or coarsened based on solution-driven criteria). These three approaches span the range from purely uniform spatial discretizations to fully dynamic spatial adaptivity, allowing us to isolate the interplay between spatial refinement and tensor compression.  

Unless otherwise specified, all calculations use SVD-based rounding~\citep{oseledets2011tensor} with a rounding tolerance of $\epsilon_{\rm TT} = 10^{-3}$.  Following~\citet{tttt}, each block employs a uniform spatial mesh and a latitude-longitude angular discretization with $N_\phi = 2N_\theta$, where quadrature points are uniformly sampled in $\cos(\theta)$ and $\phi$. 

\subsection{Behavior for Rank 1 Problems}
\label{sec:rank1_behavior}
We first analyze the limiting case in which the radiation field is globally rank one everywhere in the domain (e.g., a globally isotropic field, representative of an everywhere optically thick medium). In this limit, $r_1=r_2=1$ globally and $r_{1,b}=r_{2,b}=1$ on every block.  Consequently, there is no local rank reduction for the block-decomposed representation to exploit.  To obtain a simple analytic scaling, suppose for the moment that the grid is uniformly decomposed into $N_b$ blocks, each containing $\mathcal{N}_{\rm cells}/N_b$ cells.  Equation~(\ref{eq:global_tt_storage_general}) gives
\begin{equation}
  \mathcal{S}_{\rm \mathrm{global\text{-}TT}}
  =
  \mathcal{N}_{\rm cells}
  +
  N_\theta
  +
  N_\phi ,
  \label{eq:rank1_global_storage}
\end{equation}
while Equation~(\ref{eq:block_tt_storage_general}) gives
\begin{equation}
  \mathcal{S}_{\rm \mathrm{block\text{-}TT}}
  =
  \mathcal{N}_{\rm cells}
  +
  N_b \left( N_\theta + N_\phi \right) .
  \label{eq:rank1_block_storage}
\end{equation}
Taking $N_\phi = 2N_\theta$, the storage ratio becomes
\begin{equation}
  \mathcal{C}_d
  =
  {\mathcal{N}_{\rm cells} + 3N_\theta
  \over
  \mathcal{N}_{\rm cells} + 3N_b N_\theta} .
  \label{eq:rank1_block_ratio}
\end{equation}
Thus, when $N_\theta N_\phi \gg \mathcal{N_\mathrm{cells}}$ and $N_b \gg 1$, then $\mathcal{C}_d \propto 1/N_b$.  In other words, for a solution that is already rank one globally, domain decomposition does not expose additional low-rank structure.  Instead, the total storage grows linearly with $N_b$ due to each block maintaining its own angular representation, even though all blocks have the same rank-one structure.

\begin{figure}
        \centering
        \includegraphics[width=\textwidth]{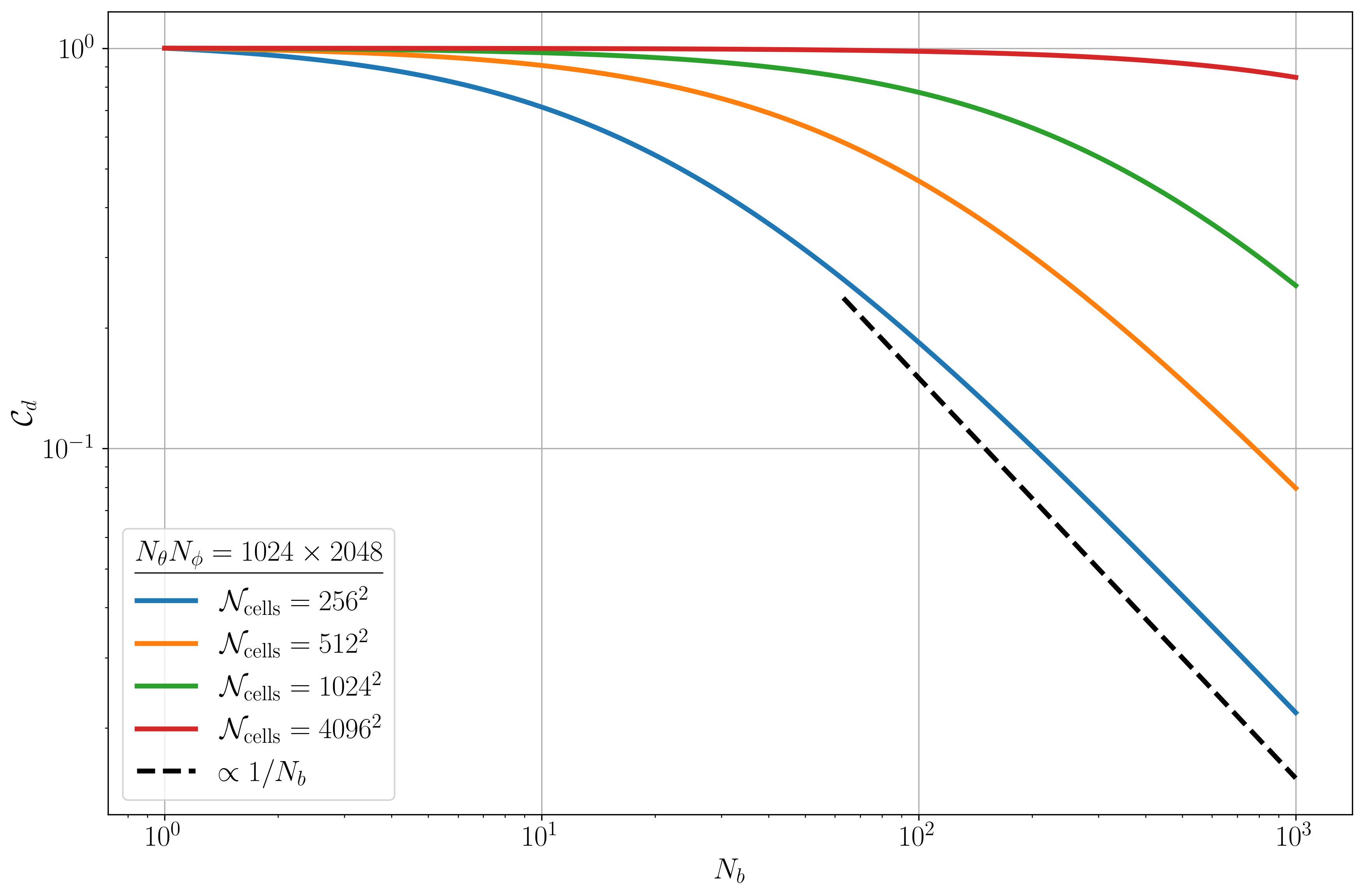}
        \caption{Compression metric $\mathcal{C}_d$ behavior as a function of number of blocks $N_b$ for rank 1 solutions with $N_\theta N_\phi = 1024 \times 2048$.  Different colors correspond to different numbers of cells $\mathcal{N_\mathrm{cells}}$.  Overplotted in black is a curve $\propto 1/N_b$, showing $\mathcal{C}_d$ analytic asymptotic predictions.}
        \label{fig:rank1}
    \end{figure}

Figure~\ref{fig:rank1} illustrates this behavior by plotting $\mathcal{C}_d$ as a function of $N_b$ for several choices of $\mathcal{N}_{\rm cells}$.  The colored curves show that the penalty associated with block decomposition is strongest when the angular storage is a large fraction of the total TT storage. For example, at the smallest grid size ($\mathcal{N}_{\rm cells} = 256^2$, blue curve), $\mathcal{C}_d$ degrades from unity to $\sim$0.03 as $N_b$ increases from 1 to 1000—a 33$\times$ penalty. In contrast, at the largest grid size ($\mathcal{N}_{\rm cells} = 4096^2$, red curve), $\mathcal{C}_d$ remains near unity until $N_b \sim 100$, after which it begins to decline. For larger $\mathcal{N}_{\rm cells}$, the spatial core dominates over a wider range of $N_b$, delaying the onset of the asymptotic degradation.  Nevertheless, once the per-block angular storage dominates the total, all curves approach the same $\mathcal{C}_d \propto 1/N_b$ scaling, shown by the black reference curve.  This confirms the simple expectation from Equation~(\ref{eq:rank1_block_ratio}): in the absence of localized rank structure, increasing the number of blocks increases total storage because the angular representation cost scales with $N_b$.

This is precisely the regime in which block adaptivity is not expected to help.  The purpose of the block-decomposed representation is to isolate localized regions of high rank so that the remainder of the domain can remain aggressively compressed.  A globally rank-one solution has no such localized rank structure to isolate.  Consequently, any observed degradation in compression for these problems is not a failure of the tensor representation, but rather the expected cost of applying a block-local tensor format to a solution that is already optimally compressed by a single global TT.

\subsection{2D Line Source with AMR}

We next consider the two-dimensional line source problem \citep[see, e.g.,][]{ls2d01, garrett13, geo_ray23}, now evolved on both uniformly decomposed and adaptively refined block-structured grids.  The problem consists of an initially isotropic pulse of radiation whose radiation energy density is formally initialized as a delta function,
\begin{equation}
    E(t=0,x,y) = \mathcal{U}_0 \delta(x,y).
    \label{eq:ls2d_delta_ic}
\end{equation}
Following \cite{geo_ray23}, the analytic solution for the evolution of this pulse in vacuum is
\begin{equation}
  E(t,x,y)
  =
  {\mathcal{U}_0 \over 2\pi}
  {\mathbf{H}\left(ct-\sqrt{x^2+y^2}\right)
  \over
  ct\sqrt{(ct)^2-(x^2+y^2)}} ,
  \label{eq:ls2d_delta_solution}
\end{equation}
where $\mathbf{H}$ is the Heaviside step function.  Because the delta-function initial condition cannot be represented on a discrete spatial grid, we instead initialize a Gaussian pulse,
\begin{equation}
  \tilde{E}(t=0,x,y)
  =
  {\mathcal{U}_0 \over 2\pi\omega^2}
  \exp\left[-{x^2+y^2 \over 2\omega^2}\right],
  \label{eq:ls2d_gaussian_ic}
\end{equation}
with width parameter $\omega$.

This problem is specifically designed to expose ray effects in $S_N$-like transport schemes \citep{brunner02}.  The pulse is initially isotropic, and therefore begins with low-rank angular structure.  As the solution evolves, however, the radiation forms an axisymmetric ring expanding outward at speed $c$.  Accurately recovering this ring requires increasing angular resolution as the radius grows.  At a final time $t_{\rm lim}$, the characteristic radius is $R_{\rm lim}\sim ct_{\rm lim}$, suggesting a critical azimuthal resolution
\begin{equation}
  N_{\phi,{\rm crit}}
  \simeq
  {2\pi R_{\rm lim} \over \Delta x}
  \label{eq:ls2d_nphi_crit}
\end{equation}
to mitigate ray effects \citep{tttt}.  Put differently, for any fixed $\Delta x$ and $N_{\rm angle}$, one can always choose a sufficiently late final time, and hence a sufficiently large propagation distance, for which ray effects and high-rank angular structure become apparent.  The line source problem therefore represents a near ``worst-case scenario'' for \texttt{TTTT}: unlike globally smooth, diffusion-like problems, the solution becomes increasingly non-separable as the pulse expands.  We now examine whether a block-local tensor representation can mitigate this difficulty by isolating high-rank structure to the localized regions where it appears, thereby achieving compression gains relative to a global TT.

\subsubsection{Block Adaptive TT Compression}

We solve this problem on a square domain with edges $[-1.28,1.28]$, choosing $c=1$, $\omega=0.03$, $\mathcal{U}_0=1$, and ${\rm CFL}\simeq 0.4$.  The specific intensity is initialized according to
\begin{equation}
  I(t=0,x,y) = J(t=0,x,y)
  =
  {c\,\tilde{E}(t=0,x,y) / [4\pi]} .
  \label{eq:ls2d_intensity_ic}
\end{equation}
To isolate the effects of domain decomposition and adaptive mesh refinement on tensor compression, we evolve this problem using two distinct grid strategies.  In the first approach, the domain is uniformly decomposed into blocks (TT-DD), with the finest spatial resolution held fixed across all calculations; increasing the number of blocks changes only the tensor decomposition, not the total number of spatial cells.  In the second approach, we employ adaptive mesh refinement (TT-AMR), with each block containing $8\times 8$ cells.  The initial Gaussian pulse is resolved at the finest AMR level in the initial condition, coarsening outside of it.  The grid is subsequently refined based on the magnitude of $J$, permitting a factor of four variation in linear resolution across the domain, thereby placing resolution near the expanding radiation front while leaving regions with negligible radiation energy comparatively coarse.

\begin{figure}
        \centering
        \includegraphics[width=\textwidth]{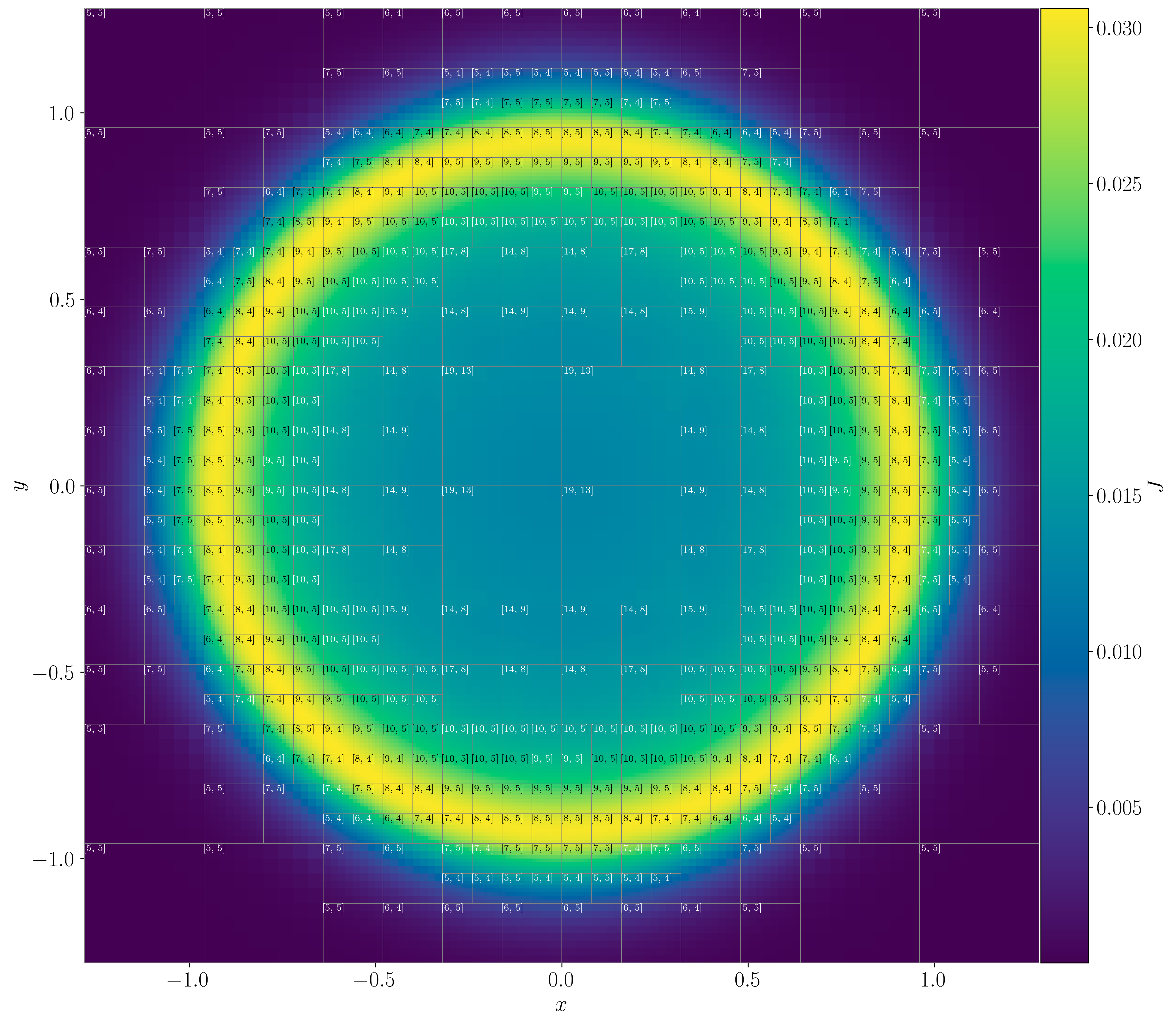}
        \caption{Final solution ($t_\mathrm{lim} = 1$) to the 2D line source problem run with AMR with an $J$ magnitude refinement criterion.  The ring is at the finest level of refinement and both its interior and exterior drop to $4\times$ coarser linear resolutions. Overplotted are per-block ranks $[r_1, r_2]$.  Note the rank increase for coarse blocks on the interior of the ring. The overplotted per-block ranks reveal the spatial heterogeneity of angular complexity: blocks on the expanding ring exhibit ranks $[r_1, r_2] \sim [8, 5]$; blocks at the domain center---where radiation energy is minimal---show elevated ranks $[r_1, r_2] \sim [19, 13]$.  The latter observation reflects the anisotropic angular structure of the radiation field near the center of the domain. Exterior blocks ahead of the wavefront maintain lower ranks.}
        \label{fig:ls2d_amr}
    \end{figure}

Figure~\ref{fig:ls2d_amr} shows a representative evolved state of the AMR line source calculation.  The refinement pattern tracks the outward-moving shell of radiation, as expected from the analytic structure of the solution.  The block boundaries are overplotted, and each block is labeled by its local TT ranks $[r_1, r_2]$.  These ranks provide a direct diagnostic of where the tensor representation must work hardest.  Blocks interior to the expanding shell require larger ranks because the local radiation field is less separable under our chosen tensor-product decomposition of spatial and angular variables.  By contrast, blocks outside the shell remain more compressible.  This spatial localization of high-rank structure is precisely the regime in which a block-local tensor representation should provide an advantage over a single global TT: rather than allowing a small subset of difficult regions to set the rank everywhere, the block-decomposed representation confines high rank to the parts of the domain that require it.

Notably, the blocks at the center of the domain also exhibit elevated ranks despite containing little radiation energy.  This occurs because our $J$-based refinement criterion targets the expanding two-dimensional shell, not the domain center, leaving the central region at coarse spatial resolution.  However, the radiation field in this region is strongly anisotropic: only angles with $\theta\approx 0$ or $\theta\approx\pi$ (i.e., nearly vertical propagation) contribute meaningfully to transport through the center.  This angular structure leads to non-separable structure that cannot be compressed effectively, even though the the total energy is low.  The result is a mismatch between spatial and angular complexity: the AMR criterion correctly identifies that high spatial resolution is unnecessary at the center, but the tensor representation still requires elevated ranks to capture the directional distribution of radiation passing through.  This highlights an important subtlety of block-local tensor compression in AMR calculations: refinement criteria based on scalar moments like $J$ may not fully capture where the solution lacks separable structure. 

\begin{figure}
        \centering
        \includegraphics[width=\textwidth]{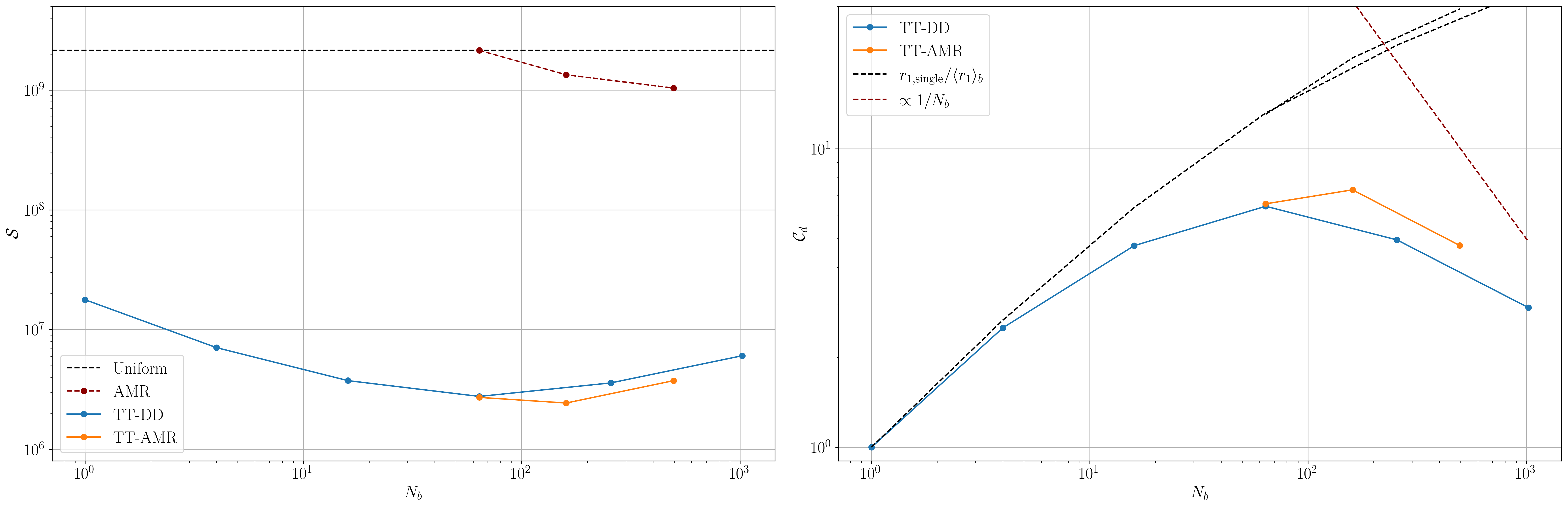}
        \caption{(Left) Storage $\mathcal{S}$ requirements for the 2D line source problem run on a TT-DD grid (blue) and a TT-AMR grid with $J$ magnitude based refinement (orange) as a function of the number of blocks each employed at the final timestep ($t_\mathrm{lim} = 1$).  $\Delta x_\mathrm{min}$ is fixed for all curves (albeit TT-AMR runs have some blocks at coarser resolutions).  Overplotted in black and dark red are the traditional storage requirements for the DD and AMR grids, respectively.  (Right) Compression metric $\mathcal{C}_d$ as a function of $N_b$ evaluated for each run.  Overplotted in black is the ratio of rank $r_1$ from the TT $N_b=1$ calculation to the mean rank across blocks $\langle r_1 \rangle_b$ for the TT-DD and TT-AMR calculations.  Overplotted in dark red is a curve $\propto 1/N_b$, showing $\mathcal{C}_d$ analytic asymptotic predictions.}
        \label{fig:ls2d_amr_compress}
    \end{figure}

While Figure~\ref{fig:ls2d_amr} illustrates where high ranks appear in the domain, a more complete assessment requires examining the total storage cost as a function of domain decomposition.  Figure~\ref{fig:ls2d_amr_compress} provides this analysis.  The left panel shows the storage requirement $\mathcal{S}$ at $t_{\rm final}=1$ as a function of the number of blocks, $N_b$, for both a tensor-train domain-decomposed grid (TT-DD; blue) and a tensor-train AMR grid (TT-AMR; orange).  For comparison, the corresponding traditional, uncompressed storage requirements are shown for the uniformly decomposed and AMR grids in black and dark red, respectively.  In the TT-DD sequence, the finest resolution $\Delta x_{\rm min}$ is fixed and the total number of spatial cells is the same as the corresponding uniform grid calculation; increasing $N_b$ changes the tensor decomposition but not the resolved physical grid.  In the TT-AMR sequence, $\Delta x_{\rm min}$ is again fixed, but not all blocks live at this finest refinement level.  The total number of spatial cells,
\begin{equation}
  \mathcal{N}_{\rm cells}^{\rm AMR}
  =
  \sum_{b=1}^{N_b}
  \mathcal{N}_{{\rm cells},b},
  \label{eq:ls2d_amr_ncells}
\end{equation}
is therefore smaller than that of the equivalent uniformly refined grid.

Consequently, the TT-AMR storage curve includes two sources of savings: fewer spatial degrees of freedom from AMR and low-rank compression from the block-local tensor representation.  Comparing TT-DD to TT-AMR therefore conflates two effects. The TT-DD calculation uses the same number of spatial cells as a uniform grid ($\mathcal{N}_{\rm cells}^{\rm uniform}$), so differences in storage relative to the uniform baseline isolate the benefit of tensor compression alone. The TT-AMR calculation, by contrast, already reduces the baseline storage requirements before tensor compression is applied, since $\mathcal{N}_{\rm cells}^{\rm AMR} < \mathcal{N}_{\rm cells}^{\rm uniform}$ by construction. However, this is precisely the point of the block-adaptive approach: in realistic calculations, tensor compression is not used in isolation from the mesh hierarchy, but rather should inherit the same spatial adaptivity that makes AMR effective.  The TT-AMR results therefore demonstrate the compounded benefit of resolving the radiation front only where necessary while also exploiting locally separable structure: the block-local representation allows ranks to vary spatially, with low ranks away from the radiation front and higher ranks only where needed.

The right panel of Figure~\ref{fig:ls2d_amr_compress} shows the corresponding compression metric,
\begin{equation}
  \mathcal{C}_d
  =
  {\mathcal{S}_{\rm \mathrm{global\text{-}TT}}^{\rm uniform} \over \mathcal{S}_{\rm \mathrm{block\text{-}TT}}},
  \label{eq:ls2d_cd}
\end{equation}
evaluated for each calculation.  Here, $\mathcal{S}_{\rm \mathrm{global\text{-}TT}}^{\rm uniform}$ denotes the storage of a single global TT representation on the equivalent uniformly refined grid.  Thus, for the TT-DD calculations, $\mathcal{C}_d$ isolates the effect of domain-local tensor compression relative to a global TT on the same grid.  For the TT-AMR calculations, $\mathcal{C}_d$ should instead be interpreted as a combined compression factor: it includes both the reduction in local TT ranks and the reduction in spatial cells produced by AMR.  We adopt this convention because it measures the quantity of practical interest: how much storage is required by the adaptive tensor calculation relative to the global, uniformly resolved tensor calculation that would otherwise be needed to achieve the same finest spatial resolution.

Unlike the rank-one limit discussed in Section~\ref{sec:rank1_behavior}, the line source develops strongly localized rank structure.  As a result, increasing $N_b$ allows blocks to maintain different local ranks adapted to their local separability.  Instead, decomposing the domain can reduce the average local rank by isolating the high-rank portion of the solution within the radiation front.  To expose this effect, we also plot the ratio
  ${r_1^{(N_b=1)} / \langle r_1\rangle_b},$
where $r_1^{(N_b=1)}$ is the first TT rank measured in the single-block calculation and $\langle r_1\rangle_b$ is the block-averaged local rank.  This ratio is not itself a storage compression factor, since the full storage also depends on $r_1r_2$, $r_2$, and the distribution of blocks across the mesh.  Nevertheless, it is a useful diagnostic for the mechanism driving $\mathcal{C}_d$: when $r_1^{(N_b=1)}/\langle r_1\rangle_b$ is large, the global TT rank is being set by localized structure that does not appear everywhere in the domain.

At $t=t_\mathrm{final}=1$, both the TT-DD and TT-AMR compression curves exhibit a clear sweet spot near $N_b\sim 100$, where $\mathcal{C}_d$ reaches $\sim$5 for TT-DD and $\sim$6 for TT-AMR. Beyond $N_b \sim 100$, $\mathcal{C}_d$ degrades to $\sim$2 at $N_b = 1000$ as the per-block angular storage overhead dominates, following the $1/N_b$ scaling predicted in \S\ref{sec:rank1_behavior}. This sweet spot reflects spatial localization of non-separable structure: high ranks are required on the expanding shell and at the domain center (where anisotropic radiation passing through drives elevated ranks despite low energy), while blocks ahead of the radiation front maintain lower rank. A block-local TT confines high ranks to the blocks that need them, whereas a global TT must carry the maximum rank everywhere. At the optimal block count ($N_b \sim 100$), the TT-DD method reduces storage to $\mathcal{S} \sim 2 \times 10^6$, 1000$\times$ smaller than the traditional uncompressed approach. The TT-AMR method achieves similar compression factors while further reducing spatial degrees of freedom in low radiation energy density regions. 

These results demonstrate that block decomposition provides substantial benefits for problems with localized angular complexity, provided the domain is not over-decomposed into excessively many blocks.  The existence of an optimal block count reflects a fundamental trade-off in the block-local tensor representation: too few blocks and the global rank is set by localized features; too many blocks and per-block angular storage overhead dominates.  Moreover, the AMR calculation illustrates an important practical point: in realistic transport problems, tensor compression and mesh adaptivity need not be competing strategies.  AMR reduces the number of spatial degrees of freedom required to resolve localized features, while the block-local TT representation reduces the rank burden away from those same features.  The combination is therefore particularly well suited to problems in which radiation occupies a small, dynamically evolving portion of the computational domain, but where accurate transport through that region still demands high angular fidelity.


\subsubsection{Rank-Based Refinement}

The previous section highlighted a subtle mismatch between spatial and angular complexity in AMR calculations: refinement criteria based on scalar moments like $J$ capture where radiation energy is concentrated, but not necessarily where the angular distribution is difficult to compress.  The elevated ranks observed at the center of the domain in Figure~\ref{fig:ls2d_amr} exemplify this issue.  Although the central region contains little radiation energy and exhibits weak spatial gradients in $J$, the strongly anisotropic angular distribution passing through it resists compression.  A $J$-based refinement criterion leaves this region coarse, forcing the tensor representation to encode the difficult angular structure on a spatially coarse grid.  This suggests a natural extension: rather than refining based on radiation energy alone, one might instead refine based on the local TT ranks themselves.  If a block exhibits high ranks---indicating that the angular distribution is difficult to compress---spatial refinement may help reduce the rank burden by better resolving the local transport structure.

We therefore repeat the AMR line source calculation with a rank-based refinement criterion.  Specifically, we monitor the local ranks $[r_1, r_2]$ in each block and refine when either exceeds a prescribed threshold (here $12$).  Conversely, blocks where both ranks fall below a lower threshold are candidates for derefinement (here, $6$).  This criterion directly couples the mesh hierarchy to the tensor representation: regions where compression is difficult automatically receive finer spatial resolution, while regions where the solution is easily compressed are allowed to coarsen.  The goal is not necessarily to reduce the total number of spatial cells---indeed, rank-based refinement may refine more aggressively than $J$-based refinement---but rather to improve the overall efficiency of the tensor representation by attempting to isolate non-separable structure.  We note that the initial condition for this problem is by definition rank-1, we therefore do not permit coarsening until after $t=0.1$.

\begin{figure}
        \centering
        \includegraphics[width=\textwidth]{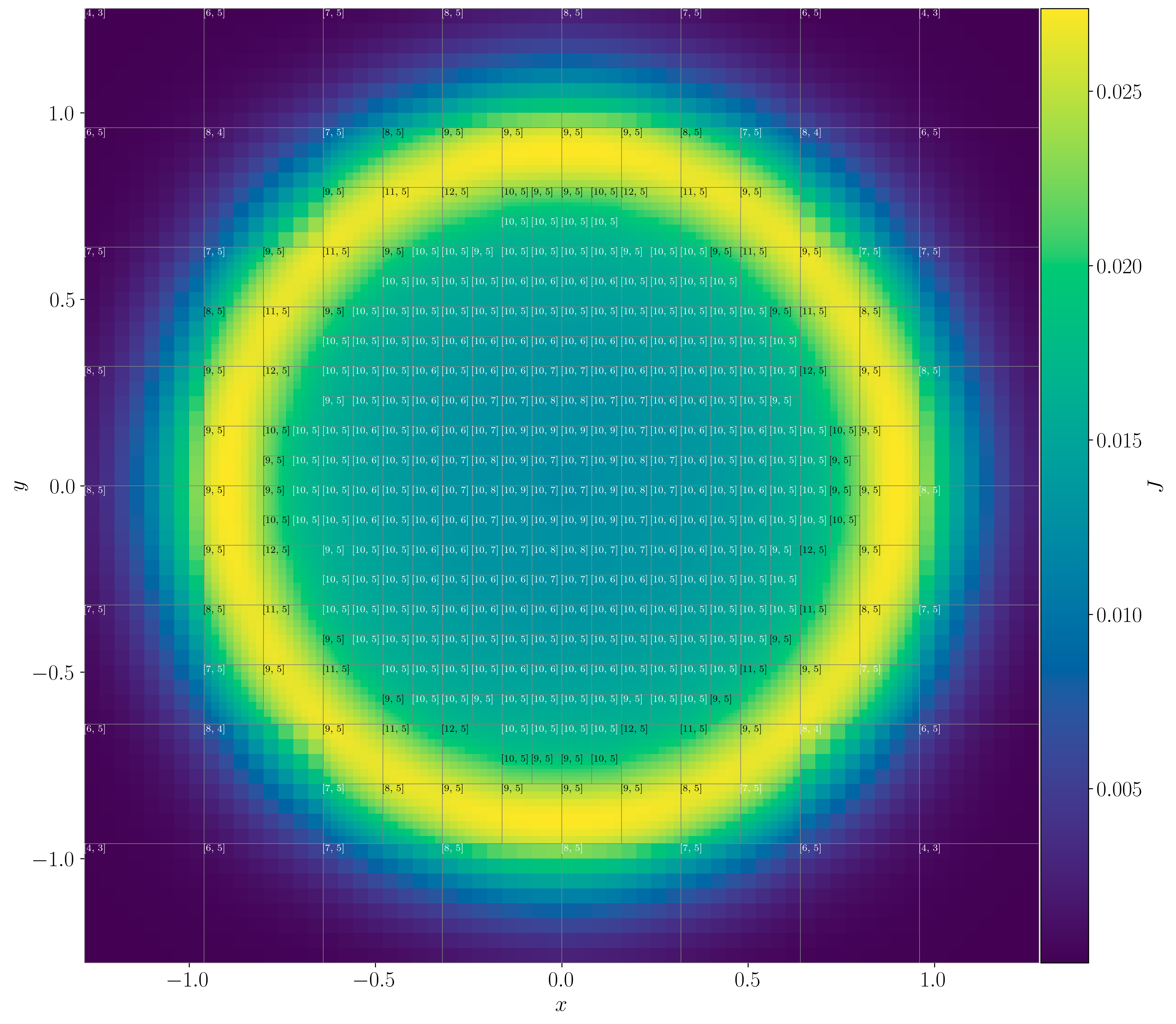}
        \caption{Final solution ($t_\mathrm{lim} = 1$) to the 2D line source problem run with AMR with a rank-based refinement criteria. Overplotted are per-block ranks $[r_1, r_2]$.  This figure should be directly compared to the $J$-based magnitude AMR criteria analogue in Figure \ref{fig:ls2d_amr}.}
        \label{fig:ls2d_criteria}
    \end{figure}

Figure~\ref{fig:ls2d_criteria} presents the final solution states ($t_{\rm lim}=1$) for the rank-based refinement strategy. The most striking difference relative to Figure~\ref{fig:ls2d_amr} is the treatment of the domain center: rank-based refinement places fine spatial resolution precisely where the $J$-based criterion left the grid coarse.  As anticipated, the elevated ranks at the center in the $J$-based calculation triggered refinement in the rank-based approach, resolving the anisotropic angular structure on a finer spatial mesh.  For our selected refinement criteria of $r_\mathrm{max} =12$ and $r_\mathrm{min} = 6$, much of the high radiation energy density ring is not resolved to the highest level of refinement.  Just behind the front is where extremal ranks are located, and where our selected criteria triggers.  Lowering $r_\mathrm{max}$ enables us to additionally refine the high-$J$ expanding ring, but we find that it also triggers refinement in regions well outside the pulse, associated with high rank structure from artificial, superluminal, numerical signals.  This will be considered further in \S{\ref{sec:sparse}}.

\begin{figure}
        \centering
        \includegraphics[width=\textwidth]{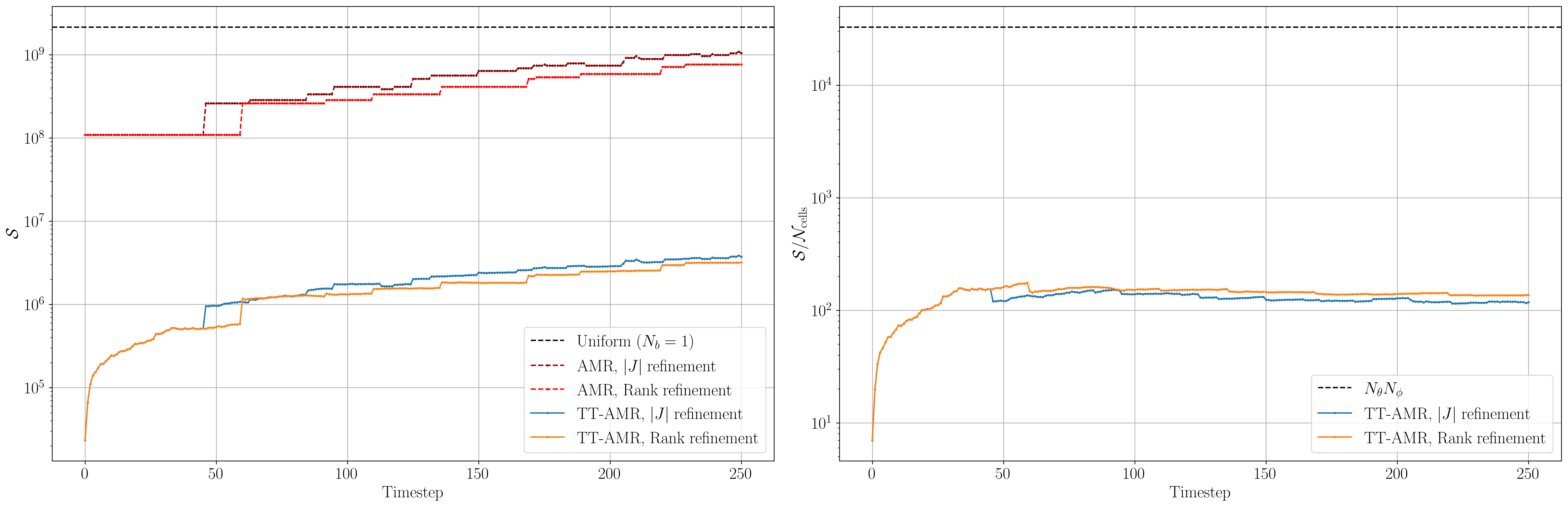}
        \caption{(Left) Storage $\mathcal{S}$ requirements for the 2D line source problem run a TT-AMR grid with (blue) $J$ magnitude based refinement and (orange) rank refinement as a function of numerical timestep.  Overplotted are the traditional storage requirements for AMR grids with corresponding refinement criteria (dark red and red, respectively).  (Right) Storage  $\mathcal{S}$ normalized by the number of cells $\mathcal{N}_\mathrm{cells}$ for each respective calculation.  Overplotted in black is the total number of angles $N_\theta N_\phi$.}
        \label{fig:ls2d_criteria_compress}
    \end{figure}

Figure~\ref{fig:ls2d_criteria_compress} compares the storage requirements for both the $J$- and rank-based refinement strategies over the course of the runs.  Both refinement strategies begin with the same initial condition, which is isotropic and therefore highly compressible; consequently, both calculations exhibit rank-1 structure at $t=0$.  As the solution evolves and the radiation pulse expands, the angular distribution becomes increasingly complex, and the tensor ranks grow accordingly.  The left panel of Figure~\ref{fig:ls2d_criteria_compress} presents $\mathcal{S}$ as a function of numerical timestep for both the $J$-based (blue) and rank-based (orange) AMR calculations.  For reference, the corresponding traditional uncompressed storage requirements are shown in dark red and red, respectively; the rank-based calculation exhibits modestly lower uncompressed storage than the $J$-based approach, since fewer blocks are employed overall.

The right panel of Figure~\ref{fig:ls2d_criteria_compress} provides further insight by normalizing the storage $\mathcal{S}$ by the total number of spatial cells $\mathcal{N}_{\rm cells}$ in each calculation.  This quantity, $\mathcal{S}/\mathcal{N}_{\rm cells}$, measures the average storage per spatial cell and is directly comparable to the total number of angles $N_\theta N_\phi$ (shown in black).  If the tensor representation achieved no compression, $\mathcal{S}/\mathcal{N}_{\rm cells}$ would equal $N_\theta N_\phi$.  Values below this reference indicate effective compression. 

This comparison highlights two competing objectives in adaptive tensor-compressed transport.  The first is to minimize spatial degrees of freedom, which $J$-based refinement does by concentrating resolution where radiation energy is significant.  The second is to exploit separable structure, which rank-based refinement addresses by refining where the tensor representation requires high ranks.  In problems where separability does not coincide with energy peaks—as in the line source—these objectives can be in tension.  Rank-based refinement prioritizes compressibility over raw cell count, accepting a modest increase in spatial resolution in exchange for improved exploitation of locally separable structure.  By monitoring tensor ranks directly, the refinement criterion adapts not only to where radiation is present, but also to where the solution exhibits less separable structure, providing a more holistic measure of computational difficulty.  In future work, hybrid refinement criteria that combine both $J$-based and rank-based triggers may offer further improvements.

\subsection{Crooked Pipe with Refinement}
Having examined the behavior of rank-1 problems (optically thick, low rank) and the 2D line source problem (optically thin, high rank) as limiting cases, we now consider a problem that exhibits both regimes concurrently.  The crooked pipe test combines optically thick walls with an optically thin interior channel, creating spatial regions with fundamentally different angular structure in the radiation field.  This heterogeneity makes it an ideal testbed for TT-AMR: the method must simultaneously handle near-isotropic diffusive regions and highly anisotropic streaming regions, adapting both spatial resolution and tensor rank as needed.

We adopt the crooked pipe configuration from \cite{southworth24} (their Figure 7).  An isotropic 0.5~keV source at the left boundary drives radiation into a complicated geometry.  Photons stream through the low-density interior, strike and heat the dense walls.  Heated walls themselves emit such that photons eventually navigate around geometric obstacles to emerge at the channel exit.  The contrasting opacity of wall and interior produces qualitatively different transport: $\sim$vacuum transport in the channel versus diffusive emission from heated boundaries.

The computational domain spans $x \in [-3.5 \; \mathrm{cm}, 3.5 \; \mathrm{cm}]$ and $y \in [-2.0 \; \mathrm{cm}, 2.0 \; \mathrm{cm}]$. Wall material has density $\rho_\mathrm{wall} = 10 \; \mathrm{g/cm^{3}}$, specific absorption opacity $\kappa_{a, \mathrm{wall}} = 200 \; \mathrm{cm^{2}/g}$, and specific heat $c_{v, \mathrm{wall}} = 10^{15} \; \mathrm{erg/g/keV}$.  The channel interior has $\rho_\mathrm{pipe} = 0.01 \; \mathrm{g/cm^{3}}$, $\kappa_{a, \mathrm{pipe}} = 20 \; \mathrm{cm^{2}/g}$, and $c_{v, \mathrm{pipe}} = c_{v, \mathrm{wall}}$.  Scattering opacity is zero throughout.  Initial conditions establish thermal equilibrium with an isotropic background at $T_{\mathrm{background}} = T_{r, \mathrm{background}} = 50 \; \mathrm{eV}$, giving $I = J = c [a_r T_{r, \mathrm{background}}^4] / (4 \pi)$.  The left boundary drives the system with an isotropic source over $-0.5 \mathrm{\;cm} \leq y \leq 0.5 \mathrm{\;cm}$ at temperature $T_{r, \mathrm{source}} = 0.5 \; \mathrm{keV}$; all other boundaries are outflow.  We integrate to $t_f = 100 \; \mathrm{ns}$ with $\mathrm{CFL} \simeq 0.4$ and the HLL flux \citep[with $\beta \simeq 5$, see][]{tttt}.

\cite{tttt} demonstrated that this problem achieves substantial compression even on uniform grids: roughly $20\times$ for $N_\mathrm{angles} = 16 \times 32$ and $275\times$ for $64 \times 128$, corresponding speedups of $\sim3\times$ and $\sim34\times$.  These gains arise because, despite the directional transport in the channel, large portions of the domain (particularly the optically thick exterior) maintain nearly isotropic fields amenable to low-rank representation.  However, the uniform-grid approach constrains all spatial locations to share the same tensor rank, forcing the entire calculation to accommodate the highest angular complexity present anywhere.  AMR offers a path beyond this limitation.

\begin{figure}
        \centering
        \includegraphics[width=\textwidth]{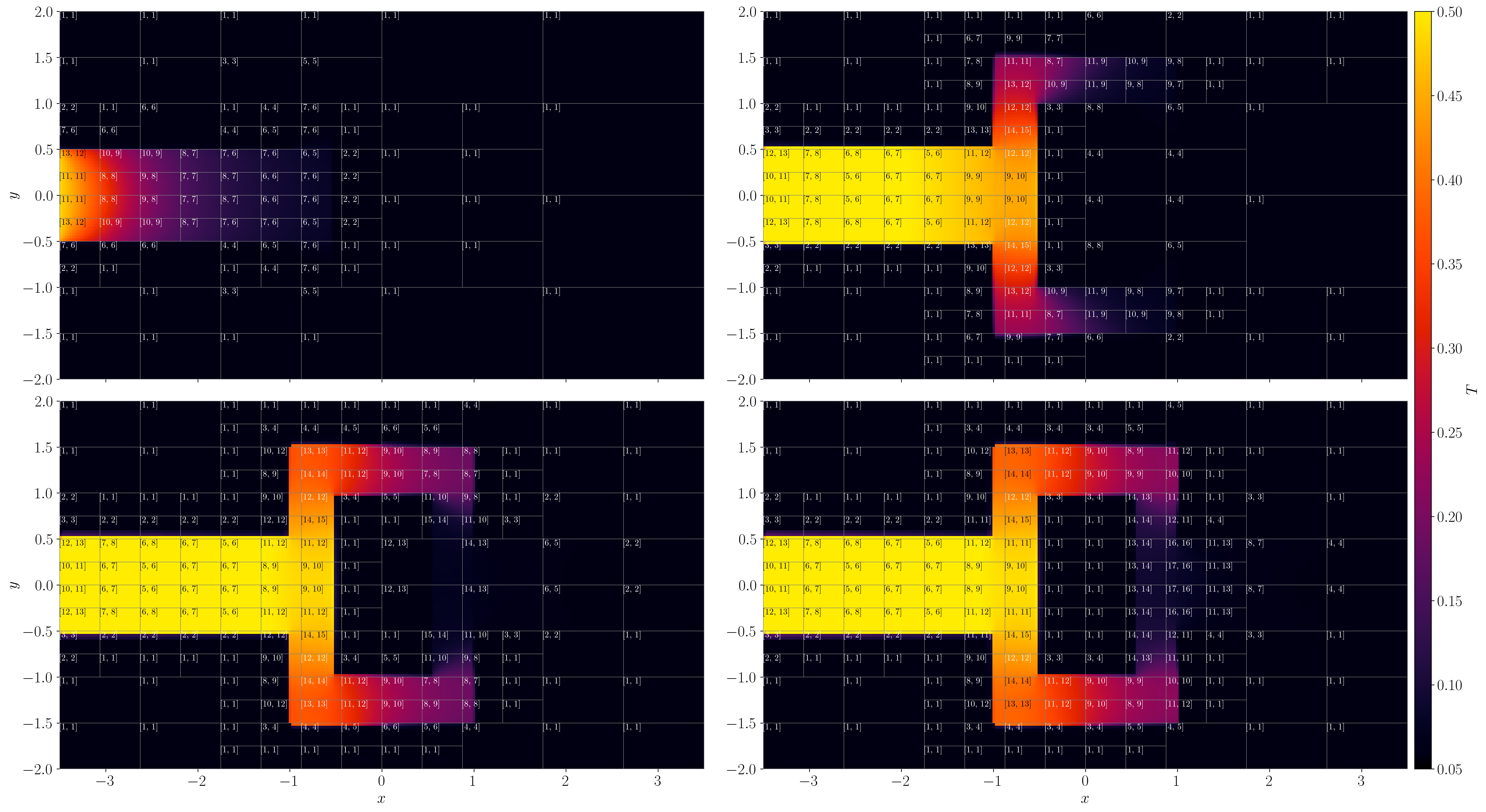}
        \caption{Crooked pipe temperature field evolution for an AMR grid with a $J$ magnitude-based refinement criterion. The four panels present snapshots at representative timesteps as radiation propagates around the obstacle.  Overplotted are per-block ranks $[r_1, r_2]$.  Note the drop-off in rank for the optically thick blocks outside the channel or where the radiation front has not yet reached inside the channel.  These four panels represent snapshots from an accompanying 12-second animation.}
        \label{fig:crooked_amr}
    \end{figure}

Figure~\ref{fig:crooked_amr} shows snapshots of the TT-AMR solution run with $J$-magnitude refinement and $N_\mathrm{angles}=32\times64$.  Spatial refinement concentrates along the active channel where radiation energy is largest, leaving the inactive exterior coarse.  Importantly, regions ahead of the radiation front where photons have not yet arrived remain at coarse resolution, with refinement dynamically tracking the advancing wavefront.  Crucially, the per-block ranks $[r_1, r_2]$ (overplotted) vary dramatically across the domain.  Blocks ahead of the radiation front maintain rank near unity, as they remain in the initial isotropic background state.  Similarly, blocks outside the channel exhibit ranks near unity---either having been heated but only seeing diffusive, isotropic thermal emission leaking from the channel walls, or maintaining their initial condition values deep inside the opaque material.  In contrast, blocks within the channel or near geometric features maintain higher ranks, as they must represent the directional source, its shadows, and the streaming propagation.  This rank heterogeneity is the key benefit of the block-local tensor factorization: each block pays only for the angular complexity it actually contains.

\begin{figure}
        \centering
        \includegraphics[width=\textwidth]{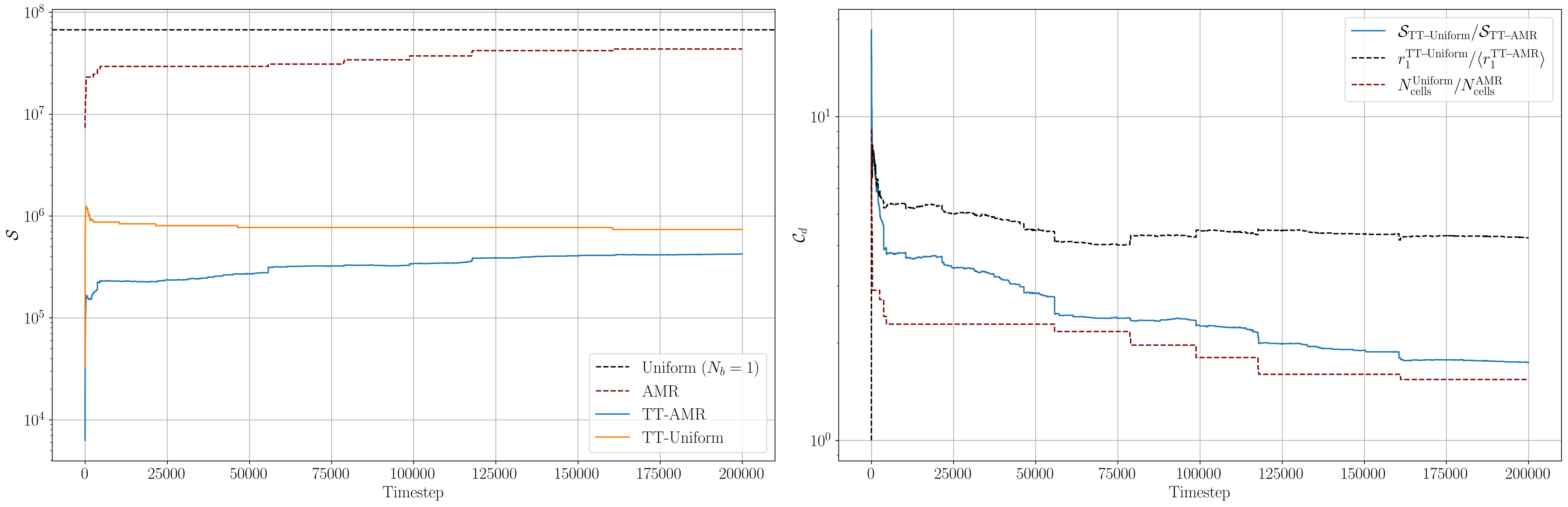}
        \caption{(Left) Storage $\mathcal{S}$ requirements for the crooked pipe problem run on (blue) a TT-uniform grid (with $N_b=1$) and (orange) a TT-AMR grid over 200k timesteps.  Overplotted are the traditional storage requirements for a corresponding uniform grid and AMR grid (black and red, respectively).  (Right) Compression metric $\mathcal{C}_d$ for the TT-AMR crooked pipe solution.  Overplotted in black is the ratio of rank $r_1$ from the TT-uniform calculation to the mean rank across blocks $\langle r_1 \rangle_b$ for the TT-AMR calculation.  Overplotted in dark red is the ratio of spatial cell counts for the uniform and AMR grid calculations (i.e., $\mathcal{N}_\mathrm{cells}^\mathrm{uniform} / \mathcal{N}_\mathrm{cells}^\mathrm{AMR}$).}
        \label{fig:crooked_compress}
    \end{figure}

Figure~\ref{fig:crooked_compress} quantifies the combined advantage.  The left panel tracks storage $\mathcal{S}$ versus time for TT-uniform (blue), TT-AMR (orange), traditional uniform (black), and traditional AMR (red).  In our previous uniform-grid study, the TT calculation exhibited a pronounced early-time storage spike when directional transport dominated, then relaxed as the field isotropized.  The TT-AMR result shows fundamentally different behavior: storage grows smoothly without a spike. In the TT-uniform calculation, the early-time storage spike reaches $\mathcal{S} \sim 10^6$—approximately 2$\times$ higher than the late-time equilibrium value—before relaxing as the field isotropizes. The TT-AMR storage increases $\sim$monotonically from $\mathcal{S} \sim 2 \times 10^5$ to $\sim 5 \times 10^5$ over the full evolution, avoiding this transient rank inflation entirely. By confining high ranks to refined blocks near the radiation front and allowing coarse exterior blocks to maintain low ranks, the method avoids the global rank increase that drives the uniform-grid storage peak.  Spatial localization of angular complexity effectively decouples different regions' rank requirements.  Additionally, by leaving unilluminated regions at coarse resolution, the method avoids paying spatial costs for portions of the domain that do not yet meaningfully contribute to the solution.

The right panel displays the compression factor $\mathcal{C}_d$ for TT-AMR, along with two diagnostic curves.  The black curve shows the ratio of uniform-grid rank to block-averaged AMR rank, $r_1 / \langle r_1 \rangle_b$, isolating the rank reduction from spatial localization.  The dark red curve shows the spatial cell count ratio, $\mathcal{N}_\mathrm{cells,uniform} / \mathcal{N}_\mathrm{cells,AMR}$, isolating the cost reduction from adaptive refinement alone.  The compression metric $\mathcal{C}_d$ captures both effects.  At early times, TT-AMR achieves a peak $\mathcal{C}_d \sim 20$. This factor reflects both fewer spatial cells (from AMR, dashed dark red) and lower average rank (from rank localization, dashed black). As the calculation progresses and radiation fills more of the domain, both factors decrease: the spatial benefit drops as more of the domain is refined, and the rank benefit similarly degrades as more blocks require high ranks. Consequently, compression decreases to $\mathcal{C}_d \sim 1.5$–$2$ at late times.

The crooked pipe demonstrates TT-AMR's effectiveness in mixed-regime transport.  Neither pure spatial adaptivity nor pure TT compression suffices: the problem requires both simultaneously.  AMR targets resolution where radiation matters spatially, while block-local tensor factorization allows each region to maintain only the angular resolution its local field structure demands.  The result is a calculation that adapts in both space and angle, matching computational cost to local solution difficulty.  The absence of a storage spike in TT-AMR, contrasting with the TT-uniform behavior, confirms that spatial localization successfully manages rank growth in time-dependent problems with evolving angular structure.


\subsection{Multigroup Stellar Irradiation with Static Refinement}

The crooked pipe demonstrated TT-AMR's ability to handle spatially heterogeneous angular complexity in a gray problem.  We now extend this capability to multigroup radiation transport, where frequency-dependent opacity introduces an additional dimension of complexity.  The multigroup stellar irradiation problem \citep{mgtttt} presents multiple radiation sources, geometric shadowing, and frequency-dependent absorption creating a solution with rich angular and spectral structure.

The computational domain spans $[-1000, 1000] \times [-1000, 1000]$ AU.  We initialize a cold circular clump of radius $R_\mathrm{clump} = 267$ AU centered at the origin, representing a dense molecular cloud core.  Two radiating stars are positioned at $(0, -500)$ AU and $(-500, 0)$ AU.  The background medium, clump, and stellar interiors have densities $\rho_\mathrm{medium} = 10^{-20}$, $\rho_\mathrm{clump} = 10^{-16}$, and $\rho_\star = 10^{-12} \; \mathrm{g/cm^3}$ respectively.  Initial temperatures are $T = T_r = 1$ K in the medium and clump, while the stellar surfaces radiate at $T = T_\star$ such that the stellar luminosities are $L = L_\odot$.  In our previous uniform-grid study \citep{mgtttt}, computational constraints limited spatial resolution.  The stellar radii were thus discretized with only a few cells.  With static mesh refinement, we concentrate resolution around the stars (and clump), allowing the stellar surfaces to be better-resolved while the distant medium remains coarse.  

We employ $N_\nu = 32$ frequency groups spanning $\nu \in [10^{12}, 10^{14}] \; \mathrm{s}^{-1}$, concentrating around the Planckian peak of the stellar emission.  Group-dependent absorption opacities vary logarithmically: $\kappa_{a,f}$ increases from $10^{-6} \; \mathrm{cm}^{2}/\mathrm{g}$ at $f=1$ to unity at $f = N_\nu/2$, then decreases back to $10^{-6}$ at $f = N_\nu$.  This frequency dependence produces qualitatively different shadowing at different photon energies.  Scattering is absent.  We use $N_\theta \times N_\phi = 128 \times 256$ angles and integrate with the Rusanov flux ($S^+ = c$) at $\mathrm{CFL} \simeq 0.4$ until a quasi-steady state is achieved.  The refined calculation employs approximately twenty billion total parameters ($\mathcal{N}_\mathrm{cells} \times N_\theta \times N_\phi \times N_\nu$), an order of magnitude fewer than the uniform-grid study but with resolution dynamically allocated where needed.

\begin{figure}
        \centering
        \includegraphics[width=\textwidth]{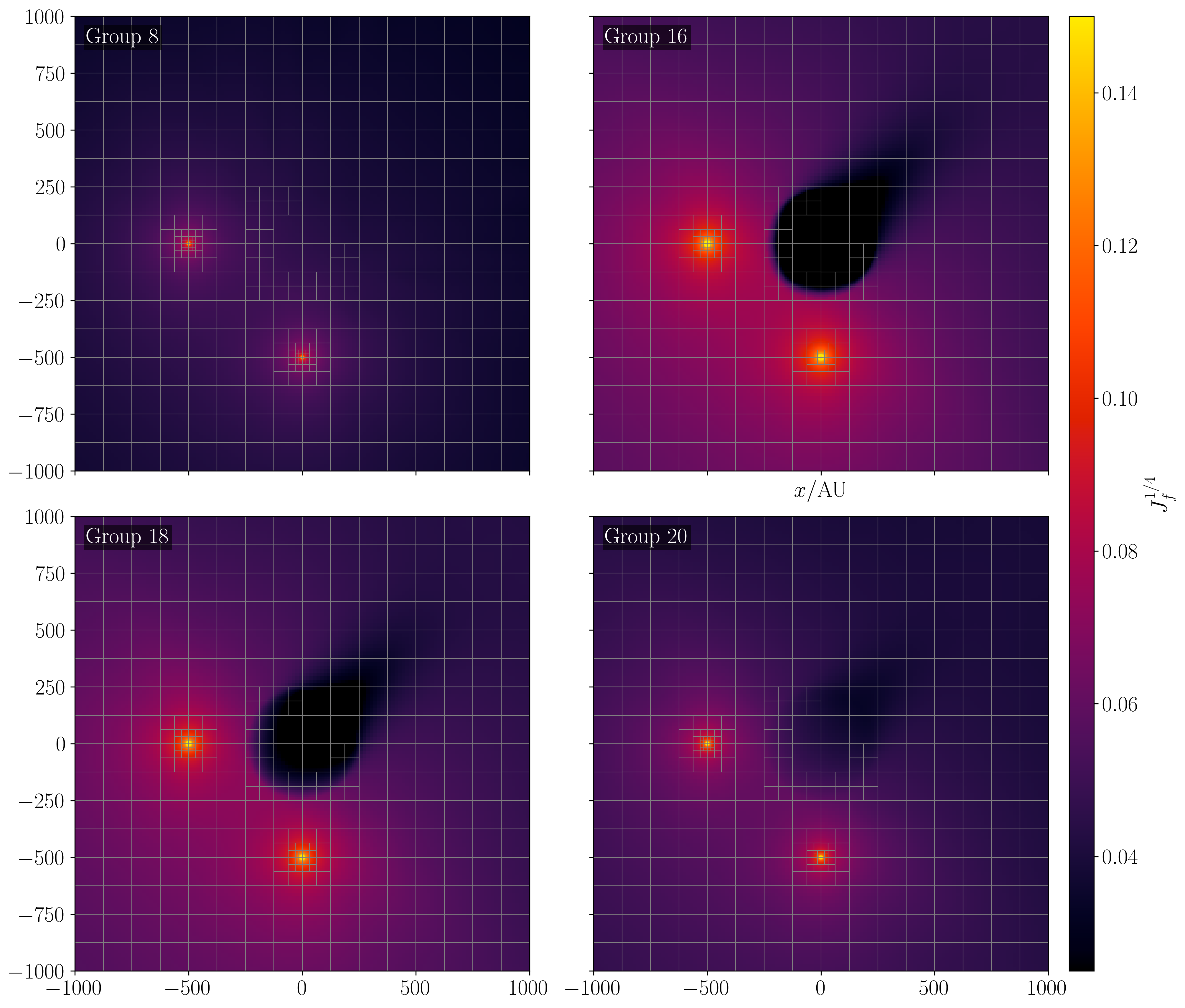}
        \caption{Multigroup stellar irradiation problem with SMR on the stars and the clump.  $\Delta x$ varies by $8\times$ across the mesh. Each panel presents solutions inside of different groups (specifically group indices 8, 16, 18, and 20).  Plotted are the mean intensities $J_f$ raised to a quarter power (to better facilitate visualization; c.f., $T_r \propto E^{1/4}$ for gray, where $T_r$ is the radiation temperature).}
        \label{fig:stellar}
    \end{figure}

Figure~\ref{fig:stellar} displays the radiation energy density $E_f^{1/4}$ for four representative groups after the system reaches quasi-equilibrium.  Group $f=16$, near peak opacity in the clump, exhibits strong shadowing—the clump effectively blocks stellar radiation, creating sharp umbral regions.  Groups farther from the opacity maximum show progressively greater penetration: $f=18$ displays partial transmission through the clump, while $f=8$ and $f=20$ at the opacity tails show nearly complete permeation.

\subsection{Discussion}
\label{sec:discussion}

\subsubsection{Conservation}

Conservation is a foundational expectation for finite-volume transport solvers.  In standard $S_N$ methods, discrete conservation follows directly from the divergence theorem: fluxes entering a cell exactly balance fluxes leaving, guaranteeing that the sum $\mathcal{E} = \sum_i E_i V_i$ remains constant (modulo source/sink terms and boundary losses) to machine precision.  In the tensor-train step-then-truncate framework, the rounding operations applied at each timestep to maintain compressed storage permit conservation only to within the rounding tolerance $\epsilon_\mathrm{TT}$. Adaptive mesh refinement introduces an additional conservation challenge at coarse-fine block interfaces, where fluxes must be carefully corrected to maintain consistency across refinement levels. We now demonstrate the interplay between these two effects: the importance of flux correction at AMR interfaces and how conservation depends on the choice of $\epsilon_\mathrm{TT}$.            

\begin{figure}
        \centering
        \includegraphics[width=\textwidth]{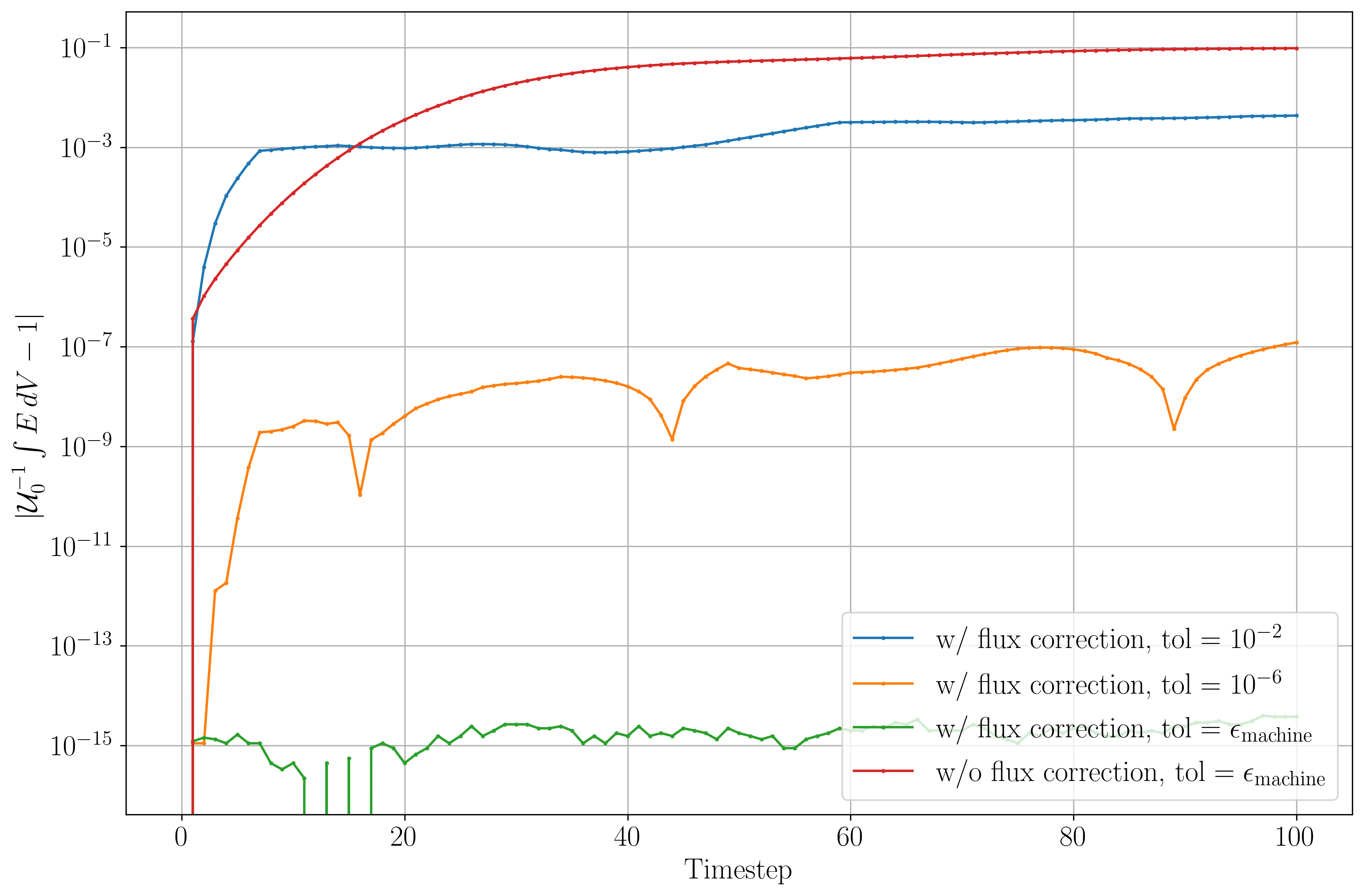}
        \caption{Conservation of total radiation energy density $(\mathcal{U}_0^{-1} \int E dV - 1)$ in the 2D line source problem with mesh refinement (SMR) as a function of numerical timestep.  Results with and without flux correction and a rounding tolerance near machine precision ($10^{-14}$) are given in blue and red, respectively.  The former conserves total energy to machine precision, whereas the latter exhibits significant deviations, underscoring the importance of flux correction. Results with flux correction but for looser rounding tolerances $10^{-6}$ and $10^{-2}$ are shown in orange and green, respectively. }
        \label{fig:ls2d_conservation}
    \end{figure}

Figure~\ref{fig:ls2d_conservation} quantifies the effects of flux correction for the same 2D line source problem from \ref{fig:ls2d_amr}, but on a TT-SMR, periodic mesh over 100 timesteps with $N_\theta \times N_\phi = 8 \times 16$.  We plot the fractional deviation in total radiation energy $\Delta \mathcal{E} \equiv \lvert \mathcal{U}_0^{-1} \int E \, dV - 1 \rvert$ versus timestep.  The blue curve shows results with flux correction enabled and $\epsilon_\mathrm{TT} =$machine epsilon.  Conservation error remains below $10^{-15}$ throughout, confirming that flux correction, when combined with tight rounding tolerances, restores discrete conservation to machine-limited accuracy.

The red curve shows the same calculation \emph{without} flux correction.  Conservation degrades rapidly, with $|\Delta \mathcal{E}|$ exceeding $10^{-1}$ by late time.  The coarse-fine interface acts as a spurious source/sink, allowing losses at resolution boundaries.

The orange ($\epsilon_\mathrm{TT} = 10^{-6}$) and green ($\epsilon_\mathrm{TT} = 10^{-2}$) curves demonstrate the impact of rounding tolerance.  Even with flux correction, looser tolerances permit larger deviations.  At $\epsilon_\mathrm{TT} = 10^{-2}$, $|\Delta \mathcal{E}| \sim 10^{-2}$—five orders of magnitude worse than the $\epsilon_\mathrm{TT} = 10^{-6}$ case.  This scaling confirms the theoretical expectation: conservation error is bounded by the chosen truncation threshold.

These results carry practical implications.  First, applications requiring strict conservation should employ tight $\epsilon_\mathrm{TT}$.  Second, users should monitor global conservation as a diagnostic—sudden drift in $\int E \, dV$ may signal inadequate rounding tolerance or improper AMR configuration. Finally, we remark that the for non-vacuum transport calculations (i.e., those coupled to a material temperature evolution equation), updating the temperature via the difference in radiation moments actually lends to global \textit{total} energy conservation --- put differently, conservation errors associated with rounding manifest as artificial heating/cooling to the fluid. 

The tolerance-limited conservation demonstrated here might motivate exploration of hybrid approaches.  Moment-based methods akin to Variable Eddington Tensor (VET) formulations \citep[e.g.,][]{jiang12,mocmc,backward_moc} could evolve angular moments with guaranteed conservation while using compressed tensor solutions of the transport equation solely for closure—the low-rank representation would provide approximate closure to an exactly conservative moment system rather than directly evolving the full angular distribution.  We do not pursue this direction further here, but note it as a potentially fruitful avenue for circumventing the conservation-compression trade-off.  Within the direct tensor transport framework adopted in this work, however, it is essential to recognize that approximate conservation is not a fundamental limitation but rather an intentional trade-off.  If storage were unlimited, one could set $\epsilon_\mathrm{TT} = \textrm{round-off}$ (i.e., retain full rank to machine precision) and recover exact conservation.  The price of compression is controlled inexactness—a bargain we willingly accept to access extreme-scale problems, provided $\epsilon_\mathrm{TT}$ is chosen judiciously and global invariants are monitored as validation metrics where possible.

\subsubsection{Sparse Blocks} \label{sec:sparse}
Combining tensor trains with domain decomposition enables a wide range of strategies for incurring errors across the computational domain.  For simulations involving blocks with widely varying magnitudes, many blocks may contribute negligibly to the solution. Rather than enforcing a prescribed rounding tolerance uniformly across the entire domain, we can restrict the error-control requirement to blocks whose Frobenius norm exceeds a threshold $F_{\rm tol}$. Blocks with $F_b \equiv \|I_b\| < F_{\rm tol}$ are assigned an effectively infinite rounding tolerance, ensuring they compress to rank one regardless of their solution structure.  We refer to this strategy as ``sparse" blocks.

\begin{figure}
    \centering
    \includegraphics[width=\textwidth]{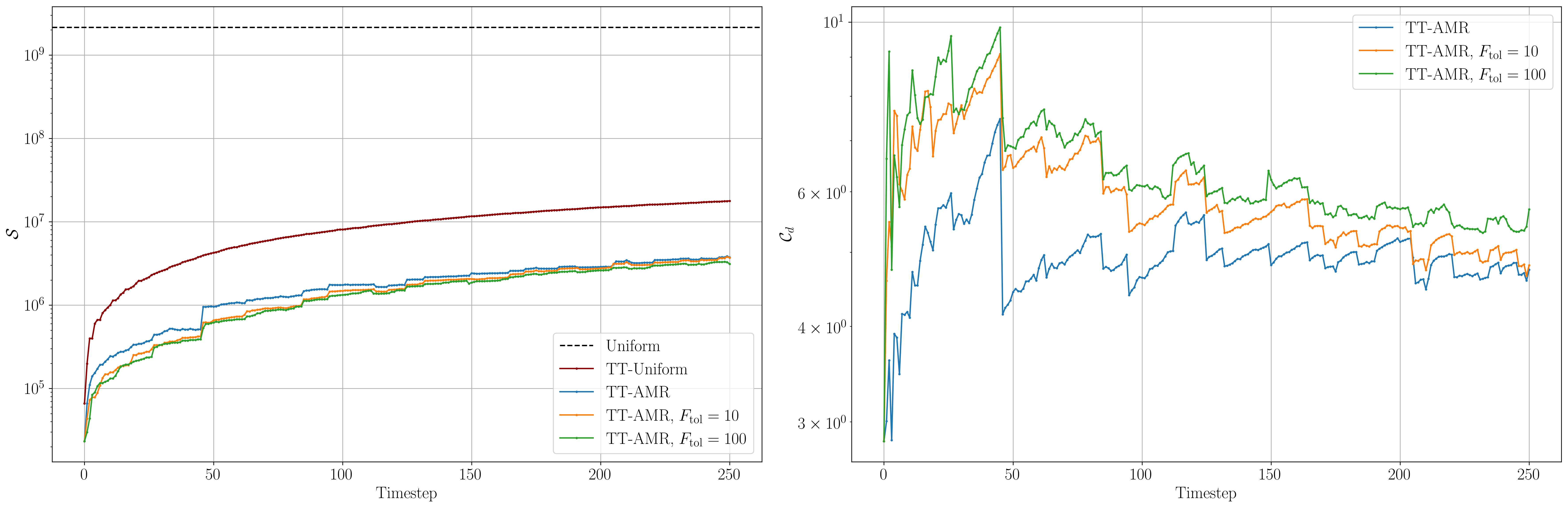}
    \caption{Sparse-block compression for the 2D line source problem with AMR. Blocks with Frobenius norm $F_b \equiv\|I_b\|_F < F_{\rm tol}$ are assigned infinite rounding tolerance, forcing rank-one compression. (Left) Storage $\mathcal{S}$ versus timestep for traditional uniform (black dashed), TT-Uniform (blue), TT-AMR without sparse treatment (orange), and TT-AMR with sparse blocks (green). Early-time compression improves by $3\times$ when many blocks remain near-zero. (Right) Compression metric $\mathcal{C}_d$ peaks at $\sim$10 for sparse blocks versus $\sim$6 without, a 60\% improvement during early propagation.}
    \label{fig:ls2d_sparse}
\end{figure}

Figure~\ref{fig:ls2d_sparse} demonstrates this sparse block strategy for the 2D line source problem with $J$-magnitude AMR. The left panel compares storage $\mathcal{S}$ over time for four cases: traditional uniform (black dashed), TT-Uniform with $N_b=1$ (blue), TT-AMR without sparse treatment (green), and TT-AMR with sparse blocks with $F_\mathrm{tol} = 10$ and $=100$ (orange and green, respectively). At early times, when much of the domain contains near-zero intensity, the sparse-block approach reduces storage from $\mathcal{S} \sim 10^6$ to $\sim 2 \times 10^5$. As the radiation fills the domain, the benefit diminishes: by timestep 100, all AMR curves $\sim$converge as fewer blocks fall below threshold. The right panel shows compression $\mathcal{C}_d$ relative to uniform TT storage. Sparse blocks achieve peak compression $\mathcal{C}_d \sim 8$ near timestep 40, compared to $\mathcal{C}_d \sim 6$ without sparse treatment.

The sparse-block strategy also eliminates spurious rank growth from numerical artifacts. First-order temporal integration introduces a superluminal numerical signal (propagating one cell per timestep) that pollutes the pre-front background with weak but non-separable structure, artificially inflating ranks in regions that should remain quiescent. Although this signal has negligible magnitude, it can drive the tensor representation to maintain elevated ranks throughout the domain. By thresholding on Frobenius norm, blocks ahead of the physical radiation front are forced to rank one, suppressing this numerical artifact without affecting the physical solution.

\begin{figure}
    \centering
    \includegraphics[width=\textwidth]{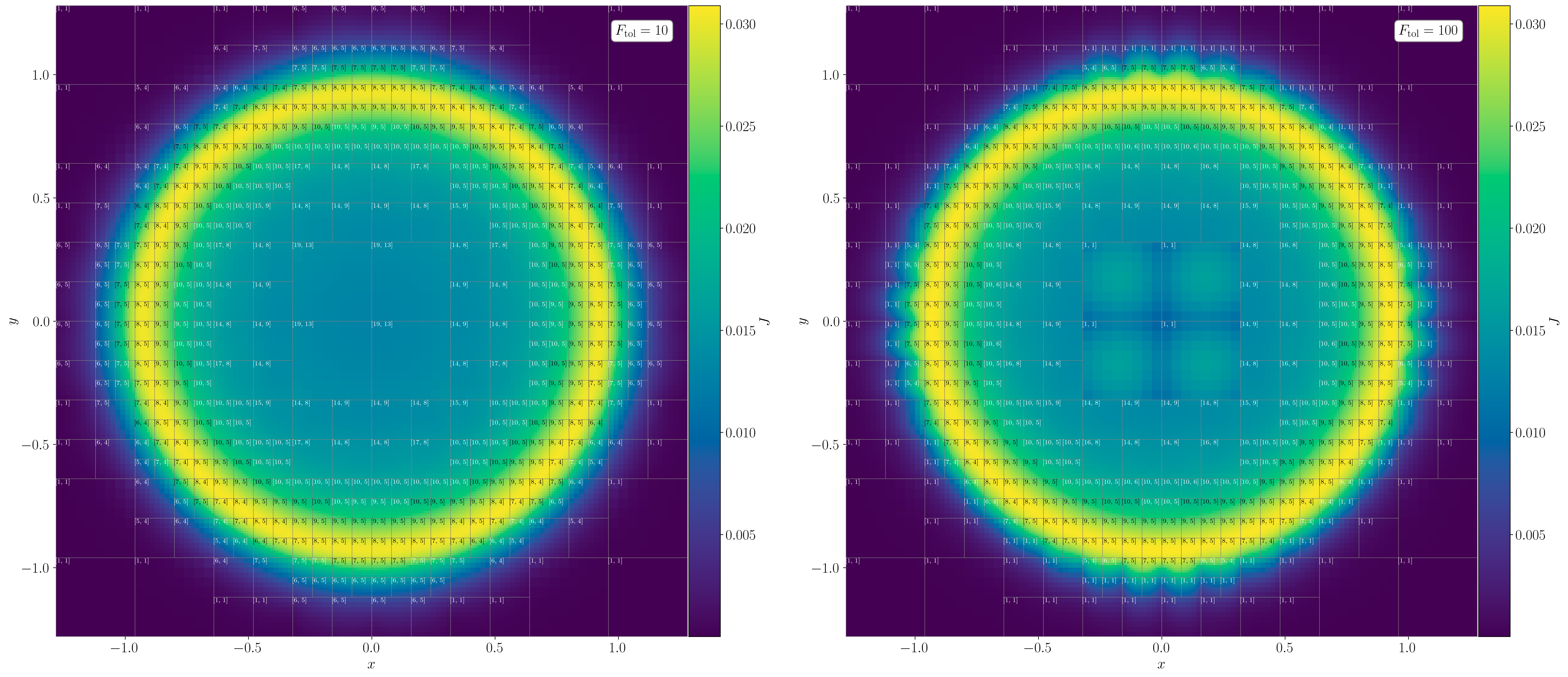}
    \caption{Final solution ($t_\mathrm{lim} = 1$) to the 2D line source problem run with AMR with an $J$ magnitude refinement criterion with the ``sparse blocks" strategy.  Two sparse tolerances $F_\mathrm{tol} = 10$ and $=1000$ are shown on the left and right panels, respectively.}
    \label{fig:ls2d_sparse_aggro}
\end{figure}

We emphasize that this approach relaxes global error control: assigning infinite tolerance to near-zero blocks provides no formal guarantee on their contribution to total error. Users must verify that $F_{\rm tol}$ is chosen conservatively relative to physically meaningful intensity scales. In Figure \ref{fig:ls2d_sparse_aggro} we present the final snapshots from the $F_\mathrm{tol}=10$ (left) and $F_\mathrm{tol}=100$ (right) runs.  The $F_\mathrm{tol}=10$ calculation is qualitatively $\sim$indistinguishable from \ref{fig:ls2d_amr}, but we can see that blocks outside the expanding front are all locked to ranks $[1,1]$.  The $F_\mathrm{tol}=100$ run demonstrates that with a very aggressive sparse tolerance, (1) the innermost blocks within the expanding ring and (2) blocks at the leading radiation front are forced to rank-1, thereby promoting large rounding errors that manifest as numerical artifacts.  

\subsubsection{Balancing Spatial and Rounding Errors}
AMR must be considered alongside tensor rounding because both contribute to the local error. In the context of AMR schemes, this implies that there is a careful balance to be reached so that AMR is not refining to an error tolerance that is well below the rounding tolerance and vice-versa. Here we summarize a practical error model to balance these effects, and we refer to Appendix~\ref{app:error_analysis} for the details. We consider the two controllable terms to be the finite-volume spatial error and the rounding error. We ignore error due to angular discretization because we do not perform adaptation in angle space, nor do we account for the interaction between the spatial error and the angular error.  For this reason,  the angular quadrature error is fixed by the chosen $S_n$ discretization and is therefore excluded from the adaptivity analysis that follows. Throughout this discussion, $\lVert \cdot \rVert$ without a subscript denotes the discrete Euclidean norm on the block-local tensor entries, while $\lVert \cdot \rVert_{L^2(\spacedomain_b \times \Omega)}$ denotes the $L^2$ norm of the corresponding piecewise-constant reconstruction over block $b$. We also use $|\nabla I|$ to denote the Euclidean magnitude of the spatial gradient.

The goal is to control a global relative tolerance by balancing the cumulative error over the full space-angle domain. For the block partition active during a specific time-step $m+1$, the squared global error is decomposed into a sum of block-local contributions,
\begin{equation}
    \left(\mathcal{E}^{m+1}\right)^2
    =
    \sum_{b=1}^N \left(\mathcal{E}^{m+1}_b\right)^2,
\end{equation}
so the adaptivity problem reduces to distributing the admissible error across blocks. Within each block, we must balance the finite-volume contribution against the rounding contribution to the error in an effort to ensure a local error satisfying
\begin{equation}
 \left(\mathcal{E}^{m+1}_b\right)^2 \leq \tau_{b,a}^2 \lVert I^{m+1}_{b,R} \rVert ^2_{L^2(\spacedomain_b \times \Omega)},
\end{equation}
where $I_{b,R}^{m+1}$ is the piecewise-constant reconstruction of the exact solution on block $b$. Note that this quantity is not computable in practice, and it is replaced by the un-rounded propagation of the approximate solution $\tilde{I}_{b,R}^{m+1}$ in the subsequent description (e.g.~\eqref{eq:refinement_criterion_summary}). Here, $\tilde{I}_{b,R}^{m+1}$ denotes the reconstruction obtained after the finite-volume update but before rounding/truncation. Under some simplifications and assumptions detailed in Appendix~\ref{app:error_analysis}, the resulting single-step blockwise error model takes the form
\begin{equation}
    \left(\mathcal{E}_b^{m+1}\right)^2 \lesssim
    2 C_{FV}^2 h_b^2 \lVert \lvert \nabla \hat{I}_{b,R}^m \rvert \rVert^2_{L^2(\spacedomain_b \times \Omega)}
    + \epsilon_b^2 \lVert \tilde{I}_{b,R}^{m+1} \rVert^2_{L^2(\spacedomain_b \times \Omega)} + 4K^2 \left(\mathcal{E}_b^{m}\right)^2,
    \label{eq:adaptivity_model}
\end{equation}
where $h_b$ is the local mesh scale, $C_{FV}$ is a constant arising from the error of the finite volume scheme, $\epsilon_b$ is the rounding tolerance, and $K$ is a stability constant of the finite volume scheme. Equation~\eqref{eq:adaptivity_model} makes the central tradeoff explicit: decreasing $h_b$ reduces the spatial discretization error at the cost of more cells and potentially higher ranks, whereas decreasing $\epsilon_b$ reduces compression error at the cost of more expensive rounding and larger retained ranks.

\paragraph{Refinement and coarsening criteria.}
To convert~\eqref{eq:adaptivity_model} into an AMR rule, we assign a fraction $\alpha \in (0,1)$ of the local error budget to the finite-volume discretization and the remaining fraction $1-\alpha$ to rounding. Moreover, we drop the term corresponding to the contribution of the accumulated error to time $m$, since it is not controllable with a time-local refinement scheme. These considerations lead to the sufficient refinement condition
\begin{equation}
    2C_{FV}^2 h_b^2 \lVert |\nabla \hat{I}_{b,R}^m| \rVert^2_{L^2(\spacedomain_b \times \Omega)}
    \le \alpha \frac{\tau_{b,a}^2}{2}\lVert \tilde{I}^{m+1}_{b,R} \rVert^2_{L^2(\spacedomain_b \times \Omega)},
    \label{eq:refinement_criterion_summary}
\end{equation}
for a chosen local error partition $\tau_{b,a}$. Blocks violating~\eqref{eq:refinement_criterion_summary} should be refined, while blocks for which the same quantity falls well below a lower threshold may be considered for coarsening. Once the spatial contribution satisfies the desired budget, the remaining error can be assigned to rounding, leading to the practical bound
\begin{equation}
    \epsilon_b^2 \le \frac{1-\alpha}{2}\tau_{b,a}^2.
\end{equation}

\paragraph{Practical implementation considerations.}
The preceding discussion isolates the transport contribution to the error budget. In coupled radiation-hydrodynamics calculations, however, refinement is often driven by variables that more directly affect the material update. In such settings, the appropriate refinement indicator depends on the full coupled system and on the discretization used for that system. For this reason, the numerical examples in Section~\ref{sec:results} also test refinement strategies that decouple the spatial indicator from the rounding tolerance, including criteria based on regions of large radiation energy density. We defer the problem-specific details to the corresponding numerical examples.

\subsubsection{Rounding Error Partitioning}

As introduced in Section \ref{sec:rounding}, the block-local formulation permits the global rounding-error budget to be distributed among blocks in different ways. The resulting tolerance-allocation strategies emphasize different aspects of the solution, such as the local intensity or block volume, and their effectiveness can therefore depend on the problem under consideration. In particular, a strategy that is advantageous in one regime may assign overly large tolerances in another and introduce visible numerical artifacts. 

\begin{figure}
    \centering
    \includegraphics[width=1.0\linewidth]{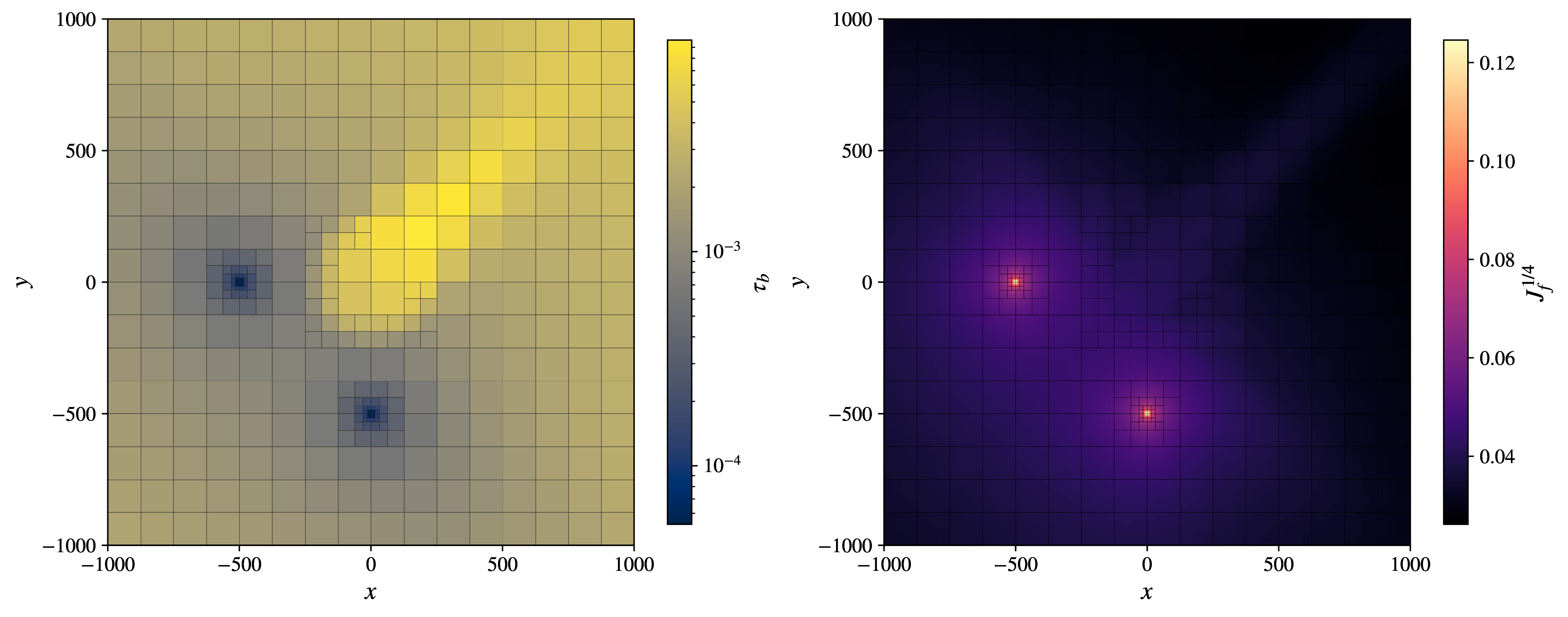}
    \caption{Effect of blockwise rounding-tolerance allocation using the volume-intensity partition scheme for the stellar irradiation problem. The volume-intensity rounding strategy with total tolerance $10^{-3}$ produces artificial illumination in the clump and its umbral region. This artifact is caused by the large local tolerances assigned to low-intensity shadowed blocks, where the blockwise tolerance $\tau_b$ is inversely proportional to the local specific-intensity norm. The corresponding tolerance map confirms that values of order $10^{-2}$ occur in the affected region.
}
    \label{fig:block_round}
\end{figure}

We illustrate this limitation using the volume-intensity rounding strategy derived in Appendix \ref{app:error_analysis} for the stellar irradiation problem in Figure \ref{fig:block_round}.  This rounding strategy invokes the volume of a block in addition to the local norm of the specific intensity to partition rounding errors across the domain.  For this case, the solution exhibits artificial illumination in the clump and in the associated umbral region. The total rounding tolerance used to generate the result was $10^{-3}$. This artifact can be understood as a consequence of assigning relatively loose local tolerances to spatial regions where the specific intensity is small. In the volume-intensity strategy, the block tolerance $\tau_b$ is inversely proportional to the norm of the specific intensity in block $b$. Therefore, the shadowed umbral region, where the radiation intensity is weak, receives a comparatively large rounding tolerance. This larger local tolerance permits errors that are visually apparent in the reconstructed radiation field.

To confirm this interpretation, we also examine the blockwise tolerances produced by the volume-intensity allocation. The tolerances in the affected umbral region are on the order of $10^{-2}$, which is sufficiently large for the corresponding rounding errors to produce the observed illumination artifacts. These artifacts can be removed either by using a uniform rounding-tolerance allocation or by reducing the total rounding tolerance from $10^{-3}$ to $10^{-4}$.

\subsection{Parallelization and Load Balancing}

Work within our block-local TT representation is also amenable to parallelization. Each block has its own TT, so expensive operations such as rounding can be done in parallel across blocks. The only synchronization required between blocks is filling ghost data and during flux correction. As a result, parallelization of block-local TT algorithms looks very similar to how e.g. a traditional AMR finite-volume hydrodynamics code would be parallelized. Nevertheless, load balancing block-local TT algorithms will be more challenging since storage and computational expense scales with bond ranks of the TTs, which are not uniform across blocks.  

\section{Conclusion}
\label{sec:conclusion}

Multiphysics simulations exhibit extreme scale separation that uniform grids cannot tractably resolve. Black hole accretion flows span from galactic scales down to the event horizon, requiring resolution of magnetohydrodynamic turbulence near the black hole while capturing large-scale inflow and radiation transport to infinity~\citep{cho24, guo25}.  Similar scale-separation challenges arise in laboratory high-energy-density physics: inertial confinement fusion must resolve micron-scale fill tubes and ablator gaps embedded in millimeter-scale targets while tracking radiation transport through surrounding hohlraums \citep{haines23}. Mesh refinement is the only viable path forward for such problems, and any transport solver targeting coupled multiphysics applications must conform to the block-structured AMR frameworks that underpin modern simulation codes.                    

This work demonstrates that tensor train methods can be successfully integrated into such frameworks.  We have developed a block-local TT representation that naturally aligns with domain decomposition, enabling TT-compressed radiation transport on adaptive octree meshes.  The key algorithmic components—prolongation, restriction, flux correction across coarse-fine interfaces, and coordinated spatial-angular rounding—have been implemented and tested across benchmarks ranging from canonical test problems to TRT scenarios.

But AMR does more than enable practical simulations—it unlocks a powerful synergy with tensor compression.  By localizing high-rank structure to blocks that need it, AMR transforms the rank distribution from a global constraint into a spatially adaptive resource.  Our test problems demonstrate this across multiple regimes: regions with anisotropic, non-separable radiation structure drive high ranks locally while diffusive regions maintain near-minimal ranks; geometric complexity requiring fine spatial resolution does not necessarily demand high angular ranks; and as physics transitions from directional transport to the diffusive regime, both spatial resolution and tensor rank can be reduced simultaneously.  Compression factors exceeding $10^3$ are achievable on problems where uniform-grid TT methods might fail due to excessive global rank requirements. Table~\ref{tab:compression_expanded} summarizes compression metrics across all test problems. 
\begin{deluxetable*}{lccccccccc}
\tablecaption{Comprehensive compression metrics for all test problems. For each case, we report statistics at minimum $\mathcal{C}_{\rm TT}$ (worst tensor compression), maximum $\mathcal{C}_{\rm TT}$ (best tensor compression), and time-averaged over all sampled iterations. Columns show: number of blocks $N_b$; total parameter count $N_{\rm params} = \mathcal{N}_{\rm cells} \times N_\theta \times N_\phi \times N_\nu$ for the traditional (uncompressed) representation; mean tensor-train ranks $\langle r_1 \rangle$ and $\langle r_2 \rangle$; and three compression factors. $\mathcal{C}_{\rm AMR} = \mathcal{N}_{\rm cells}^{\rm uniform} / \mathcal{N}_{\rm cells}^{\rm AMR}$ quantifies spatial compression from adaptive mesh refinement relative to a uniform grid at the finest resolution. $\mathcal{C}_{\rm TT} = N_{\rm params} / S_{\rm TT}$ quantifies compression from tensor-train decomposition, where $S_{\rm TT}$ is the actual storage cost of the TT representation. $\mathcal{C}_{\rm total} = (\mathcal{N}_{\rm cells}^{\rm uniform} \times N_\theta \times N_\phi \times N_\nu) / S_{\rm TT}$ combines both AMR and TT compression relative to a traditional uniform-grid representation. Dashes indicate no spatial refinement (pure domain decomposition or uniform grids). \label{tab:compression_expanded}}
\tablehead{
Method &
Metric &
$N_b$ &
$N_{\rm params}$ &
$\langle r_1 \rangle$ &
$\langle r_2 \rangle$ &
$\mathcal{C}_{\rm AMR}$ &
$\mathcal{C}_{\rm TT}$ &
$\mathcal{C}_{\rm total}$
}
\startdata
\multicolumn{9}{c}{2D Line Source ($N_\theta = 128$, $N_\phi = 256$, $N_\nu = 1$)} \\
\hline
TT-AMR (rank-based) & Min $\mathcal{C}_{\rm TT}$ & 52 & $1.1 \times 10^{8}$ & 9 & 7 & 20 & 186 & 3,670 \\
 & Max $\mathcal{C}_{\rm TT}$ & 52 & $1.1 \times 10^{8}$ & 1 & 1 & 20 & 4,680 & 92,200 \\
 & Mean & 179 & $3.8 \times 10^{8}$ & 8 & 6 & 9 & 270 & 2,840 \\
\hline
TT-AMR ($J$-magnitude) & Min $\mathcal{C}_{\rm TT}$ & 52 & $1.1 \times 10^{8}$ & 8 & 7 & 20 & 209 & 4,120 \\
 & Max $\mathcal{C}_{\rm TT}$ & 52 & $1.1 \times 10^{8}$ & 1 & 1 & 20 & 4,680 & 92,200 \\
 & Mean & 244 & $5.1 \times 10^{8}$ & 8 & 5 & 7 & 296 & 2,610 \\
\hline
TT-DD ($N_b=1$) & Min $\mathcal{C}_{\rm TT}$ & 1 & $2.1 \times 10^{9}$ & 248 & 47 & -- & 121 & -- \\
 & Max $\mathcal{C}_{\rm TT}$ & 1 & $2.1 \times 10^{9}$ & 1 & 1 & -- & 32,600 & -- \\
 & Mean & 1 & $2.1 \times 10^{9}$ & 136 & 32 & -- & 617 & -- \\
\hline
TT-DD ($N_b=4$) & Min $\mathcal{C}_{\rm TT}$ & 4 & $2.1 \times 10^{9}$ & 93 & 20 & -- & 304 & -- \\
 & Max $\mathcal{C}_{\rm TT}$ & 4 & $2.1 \times 10^{9}$ & 1 & 1 & -- & 32,000 & -- \\
 & Mean & 4 & $2.1 \times 10^{9}$ & 53 & 16 & -- & 976 & -- \\
\hline
TT-DD ($N_b=16$) & Min $\mathcal{C}_{\rm TT}$ & 16 & $2.1 \times 10^{9}$ & 39 & 12 & -- & 573 & -- \\
 & Max $\mathcal{C}_{\rm TT}$ & 16 & $2.1 \times 10^{9}$ & 1 & 1 & -- & 30,000 & -- \\
 & Mean & 16 & $2.1 \times 10^{9}$ & 22 & 12 & -- & 1,780 & -- \\
\hline
TT-DD ($N_b=64$) & Min $\mathcal{C}_{\rm TT}$ & 64 & $2.1 \times 10^{9}$ & 19 & 11 & -- & 662 & -- \\
 & Max $\mathcal{C}_{\rm TT}$ & 64 & $2.1 \times 10^{9}$ & 1 & 1 & -- & 23,800 & -- \\
 & Mean & 64 & $2.1 \times 10^{9}$ & 12 & 8 & -- & 2,160 & -- \\
\hline
TT-DD ($N_b=256$) & Min $\mathcal{C}_{\rm TT}$ & 256 & $2.1 \times 10^{9}$ & 10 & 7 & -- & 540 & -- \\
 & Max $\mathcal{C}_{\rm TT}$ & 256 & $2.1 \times 10^{9}$ & 1 & 1 & -- & 13,100 & -- \\
 & Mean & 256 & $2.1 \times 10^{9}$ & 7 & 5 & -- & 1,410 & -- \\
\hline
TT-DD ($N_b=1024$) & Min $\mathcal{C}_{\rm TT}$ & 1024 & $2.1 \times 10^{9}$ & 7 & 4 & -- & 354 & -- \\
 & Max $\mathcal{C}_{\rm TT}$ & 1024 & $2.1 \times 10^{9}$ & 1 & 1 & -- & 4,680 & -- \\
 & Mean & 1024 & $2.1 \times 10^{9}$ & 4 & 3 & -- & 837 & -- \\
\hline
\multicolumn{9}{c}{Crooked Pipe ($N_\theta = 32$, $N_\phi = 64$, $N_\nu = 1$)} \\
\hline
TT-AMR & Min $\mathcal{C}_{\rm TT}$ & 118 & $3.1 \times 10^{7}$ & 6 & 6 & 2 & 96 & 208 \\
 & Max $\mathcal{C}_{\rm TT}$ & 28 & $7.3 \times 10^{6}$ & 1 & 1 & 9 & 1,170 & 10,700 \\
 & Mean & 138 & $3.6 \times 10^{7}$ & 5 & 5 & 2 & 109 & 211 \\
\hline
TT-Uniform & Min $\mathcal{C}_{\rm TT}$ & 1 & $6.7 \times 10^{7}$ & 37 & 20 & -- & 54 & -- \\
 & Max $\mathcal{C}_{\rm TT}$ & 1 & $6.7 \times 10^{7}$ & 1 & 1 & -- & 2,040 & -- \\
 & Mean & 1 & $6.7 \times 10^{7}$ & 23 & 21 & -- & 86 & -- \\
\hline
\multicolumn{9}{c}{Stellar Irradiation ($N_\theta = 128$, $N_\phi = 256$, $N_\nu = 32$)} \\
\hline
TT-SMR & Min $\mathcal{C}_{\rm TT}$ & 355 & $2.4 \times 10^{10}$ & 8 & 7 & 46 & 2,570 & 119,000 \\
 & Max $\mathcal{C}_{\rm TT}$ & 355 & $2.4 \times 10^{10}$ & 1 & 1 & 46 & 27,600 & 1,270,000 \\
 & Mean & 355 & $2.4 \times 10^{10}$ & 7 & 6 & 46 & 3,000 & 139,000 \\
\enddata
\tablecomments{Storage and compression metrics exclude the cost of ghost cells used for inter-block communication. All metrics are computed using post-rounding storage requirements rather than pre-rounded tensor sizes; see \citet{mgtttt} Appendix A for a detailed discussion of these reporting conventions and their impact on compression interpretations.}
\end{deluxetable*}

These benefits come with important caveats.  Block count matters: too few blocks undermine localization of high rank structure; too many inflate per-block tensor overhead.  Refinement criteria must couple spatial discretization error to compression error, balancing grid resolution against tensor rank.  And while structured octree meshes provide a natural starting point, we have not yet presented globally implicit solutions to TRT with tensors.  

Despite these complexities, the results are compelling.  Tensors permit much lower memory footprints, hence combatting the curse of dimensionality in (multigroup) thermal radiation transport.  The framework integrates naturally with existing AMR infrastructure, positioning tensor methods as a drop-in enhancement for production multiphysics codes.  Tensor methods and adaptive mesh refinement are not merely compatible—they are mutually reinforcing.  Together, they provide a scalable path toward high-fidelity multiphysics simulation at resolutions previously deemed intractable. 

\clearpage
\appendix
\section{Rounding and error analysis}\label{app:error_analysis}
    
As discussed in Section~\ref{sec:tensors}, rounding is the key mechanism that prevents runaway rank growth. In the AMR setting, it must be balanced against the spatial truncation error introduced by the finite-volume discretization. In this appendix, we develop a practical error model for how these two effects contribute to the total error. We restrict attention to piecewise-constant reconstruction; extending the analysis to higher-order nonlinear schemes would require a more detailed treatment of how rounding interacts with nonlinear reconstruction.

\subsection{Error analysis}
There are two critical errors that must be balanced: finite volume spatial error and rounding error. The quadrature error due to the angular discretizations is neglected because that discretization is fixed and cannot be mitigated. As a result, we define the reference solution with respect to which our analysis is performed $I^m(x,\omega)$, where $\omega = (\theta, \phi)$, as the exact continuous-space solution to the discrete-ordinate $(S_n)$ equations. We also let $I_R^m(x,\omega)$ denote its piecewise-constant cell-average reconstruction on the active block partition. Consequently, the static angular quadrature error is decoupled from our analysis, and our error strictly measures the errors introduced by the finite volume grid and tensor compression.

Consider the error incurred over a single time step. We begin with a discrete solution $\hat{I}^m$ and its reconstruction $\hat{I}_R^m(x, \omega)$. The step-and-truncate solver first applies the discrete finite-volume update to produce $\tilde{I}^{m+1}$, with reconstruction $\tilde{I}_R^{m+1}(x, \omega)$. It then rounds $\tilde{I}^{m+1}$ to a new approximation $\hat{I}^{m+1}$, with reconstruction $\hat{I}_{R}^{m+1}(x, \omega)$. This rounding is performed blockwise, and the rounding algorithm guarantees an accuracy tolerance $\epsilon_b$
\begin{equation}
    \lVert \tilde{I}_b^{m+1} - \hat{I}_b^{m+1} \rVert \leq \epsilon_b \lVert \tilde{I}_b^{m+1} \rVert,
\end{equation}
where we have noted the solutions restricted to a specific block.

Our goal is to control a global error tolerance $\tau$ to achieve 
\begin{equation}
    \left(\mathcal{E}^{m+1}\right)^2 \equiv \int_\Omega \int_\mathcal{X} \left(I_R^{m+1}(x, \omega) - \hat{I}_R^{m+1}(x, \omega)\right)^2 dx\, d\omega \leq \tau^2 \int_\Omega \int_\mathcal{X} \left(I_R^{m+1}(x,\omega)\right)^2 dx\, d\omega,
    \label{eq:global_error_tol}
\end{equation}
where $d \omega = \cos(\phi)\, d\theta\, d\phi$ in our angular coordinates.

For the block partition active during this step, the global error decomposes into a sum of block-local errors,
    \begin{equation}
    \left(\mathcal{E}^{m+1}\right)^2 = \sum_{b=1}^N\left(\int_\Omega \int_{\mathcal{X}_b} \left(I_{b,R}^{m+1}(x, \omega) - \hat{I}_{b,R}^{m+1}(x, \omega)\right)^2 dx\, d\omega \right)
    =  \sum_{b=1}^N \left(\mathcal{E}^{m+1}_b\right)^2,
\end{equation}
where $\mathcal{E}_b^{m+1}$ denotes the block-local error. We now separate the finite-volume and rounding contributions by introducing the intermediate intensity
\begin{align}
    \left(\mathcal{E}_b^{m+1}\right)^2 &= \int_\Omega \int_{\mathcal{X}_b} \left(\left(I_{b,R}^{m+1}(x,\omega) - \tilde{I}^{m+1}_{b,R}(x, \omega)\right) + \left(\tilde{I}^{m+1}_{b,R}(x, \omega) - \hat{I}^{m+1}_{b,R}(x,\omega)\right)\right)^2 dx\, d\omega \\
    &\leq 2\left[ \int_\Omega \int_{\mathcal{X}_b} \left(I_{b,R}^{m+1}(x,\omega) - \tilde{I}^{m+1}_{b,R}(x, \omega)\right)^2  dx\, d\omega + \int_\Omega \int_{\mathcal{X}_b}  \left(\tilde{I}^{m+1}_{b,R}(x, \omega) - \hat{I}^{m+1}_{b,R}(x,\omega)\right)^2 dx\, d\omega \right]
    \label{eq:error_split}
\end{align}
The first term represents the error introduced by the spatial discretization, while the second term is the error due to rounding.

The second term is easier to bound using the rounding tolerance. We first need to move from a discrete bound on the cell-centered values to the reconstruction error. Because we are doing piecewise-constant reconstruction in each block, we can relate the two errors based on the cell volume $\Delta V_b$ and the angular quadrature weights. Under the simplifying assumption that the quadrature is uniform in solid angle, this relation takes the form
\begin{equation}
\int_\Omega \int_{\mathcal{X}_b}  \left(\tilde{I}^{m+1}_{b,R}(x, \omega) - \hat{I}^{m+1}_{b,R}(x,\omega)  \right)^2 dx\, d\omega  = \Delta V_b \Delta \Omega  \lVert \tilde{I}_b^{m+1} - \hat{I}_b^{m+1} \rVert^2 ,
\end{equation}
where the factor $\Delta \Omega$ arises from a quadrature rule that is uniform in solid angle. In other cases, it is replaced by the corresponding weighted discrete inner product.
The normalization term has the analogous form
\begin{equation}
\int_\Omega \int_{\mathcal{X}_b}  \left(\tilde{I}^{m+1}_{b,R}(x, \omega)\right)^2   dx\, d\omega  = \Delta V_b \Delta \Omega  \lVert \tilde{I}_b^{m+1} \rVert^2.
\end{equation}
Thus, rounding the discrete tensor also ensures the following error bound 
\begin{equation}
    \int_\Omega \int_{\mathcal{X}_b}  \left(\tilde{I}^{m+1}_{b,R}(x, \omega) - \hat{I}^{m+1}_{b,R}(x,\omega)  \right)^2 dx\, d\omega \leq \epsilon_b^2 \int_\Omega \int_{\mathcal{X}_b}  \left(\tilde{I}^{m+1}_{b,R}(x, \omega)\right)^2   dx\, d\omega.
    \label{eq:term2_error}
\end{equation}

To estimate the first term in~\eqref{eq:error_split}, we appeal to standard consistency and stability properties of a first-order finite-volume method. To begin with, we split this term into two components. Let $\Pi_b$ denote the cell-average projection onto the finite-volume space on block $b$, so that $I_{b,R}^{m} = (\Pi_b I^m)_R$. Further, let $\mathcal{P}^m$ be the exact propagation operator, satisfying $I^{m+1} = \mathcal{P}^m[I^m]$, and let $\mathcal{N}^m_b$ be an effective block-local finite-volume propagation operator, including ghost-cell communication and flux correction on the active mesh, satisfying $\tilde{I}_b^{m+1} = \mathcal{N}_b^m[\hat{I}^m_b]$. Using these operators, we decompose the spatial error in the domain $\spacedomain_b$ as 
\begin{align}
I_{b,R}^{m+1} - \tilde{I}^{m+1}_{b,R} &= (\Pi_b \mathcal{P}^m[I^{m}])_R - (\mathcal{N}^m_b[\hat{I}_{b}^m])_R \\ 
&= \left((\Pi_b \mathcal{P}^m[I^{m}])_R - (\mathcal{N}^m_b[\Pi_b I^{m}])_R\right) + \left((\mathcal{N}_b^m[\Pi_b I^m])_R - (\mathcal{N}_b^m[\hat{I}_b^m])_R\right),
\end{align}
where the first term corresponds to the local truncation error incurred by the numerical scheme had we started the time-step with the true answer, and the second error corresponds to the propagation error associated with starting from the approximate solution at the $m$-th timestep.

For sufficiently smooth solutions, we model the local truncation error of the first-order finite-volume discretization in the standard form \citep{leveque2002finitevolume}
\begin{equation}
   \int_\Omega \int_{\spacedomain_b} \left((\Pi_b \mathcal{P}^m[I^{m}])_R(x,\omega) - (\mathcal{N}^m_b[\Pi_b I^{m}])_R(x,\omega)\right)^2 dx\, d\omega \leq C_{FV}^2 h_b^2 \int_\Omega \int_{\spacedomain_b} |\nabla I^m|^2(x,\omega) dx\, d\omega,  \label{eq:FVE}
\end{equation}
where $C_{FV}$ is a constant depending on the scheme and dimensionality. We further assume that the finite-volume update is $L_2$-stable with Lipschitz constant $K$, so that the second term can be bounded as
\begin{equation}
   \int_\Omega \int_{\spacedomain_b} \left((\mathcal{N}_b^m[\Pi_b I^{m}])_R(x,\omega) - (\mathcal{N}^m_b[\hat{I}^{m}_b])_R(x,\omega)\right)^2 dx\, d\omega \leq K^2 \int_\Omega \int_{\spacedomain_b} \left(I_{b,R}^m(x,\omega) - \hat{I}^m_{b,R}(x,\omega)\right)^2 dx\, d\omega 
   = K^2 \left(\mathcal{E}_b^m\right)^2    \label{eq:KE}
\end{equation}
which yields the recursive dependence on the previous-step error. Combining these terms and using the triangle inequality gives
\begin{equation}
     \int_\Omega \int_{\spacedomain_b} \left(I_{b,R}^{m+1}(x,\omega) - \tilde{I}^{m+1}_{b,R}(x, \omega)\right)^2  dx\, d\omega \leq 2\left[ C_{FV}^2 h_b^2 \int_\Omega \int_{\spacedomain_b} |\nabla I^m|^2(x,\omega) dx\, d\omega + K^2 \left(\mathcal{E}_b^m\right)^2 \right]
     \label{eq:term1_error}
\end{equation}
Using ~\eqref{eq:term2_error} and~\eqref{eq:term1_error} within ~\eqref{eq:error_split} yields 
\begin{equation}
\label{eq:err_integral}
    \left(\mathcal{E}_b^{m+1}\right)^2 \leq 4\left[C_{FV}^2 h_b^2\int_\Omega \int_{\spacedomain_b} |\nabla I^m|^2(x,\omega) dx d\omega  + K^2\left(\mathcal{E}_b^m\right)^2    \right]  + 2 \epsilon_b^2 \int_\Omega \int_{\mathcal{X}_b}  \left(\tilde{I}^{m+1}_{b,R}(x, \omega)\right)^2 dx d\omega,
\end{equation}
where the first term contributes to the local finite volume error, the second term involves the recursive error from previous time-steps, and the last term corresponds to the rounding error. If we use $\lVert \cdot \rVert_{L^2(\spacedomain\times\Omega)}$ to represent the above integrals, we obtain the following error bound
\begin{equation}
   \left(\mathcal{E}^{m+1}\right)^2  \leq 2 \sum_{b}\left( 2 C_{FV}^2 h_b^2 \lVert \lvert \nabla I^m \rvert\rVert^2_{L^2(\spacedomain_b \times \Omega)} 
   + \epsilon_b^2 \lVert \tilde{I}_{b,R}^{m+1}  \rVert^2_{L^2(\spacedomain_b \times \Omega)}
   \right) + 4K^2 \sum_{b} \left(\mathcal{E}^m_b\right)^2 \label{eq:error_bound}
\end{equation}
where the first summation on the right-hand-side corresponds to the new error incurred on the current time-step and the second summation corresponds to the accumulated error. 

\subsection{Rounding criteria}

The error bound~\eqref{eq:error_bound} provides a heuristic model for balancing mesh adaptivity and rounding. We now use it to motivate practical choices that couple mesh refinement and rounding. In coupled radiation-hydrodynamics calculations, mesh refinement is often driven by additional physics variables rather than by the transport solve alone. Nevertheless, the model clarifies how spatial discretization and rounding contribute to the local error budget and therefore provides guidance for setting rounding tolerances in such coupled calculations.

To obtain computable relations for the tradeoff between errors, we make two simplifying assumptions. First, the time-stepping scheme is fixed, so $K$ is not adjustable. Second, we partition error equally across time steps, so the target is the same local error at each step. Under these assumptions, we drop the $4K^2$ term because it is not directly controllable, and instead partition $\tau^2 \lVert I_R^{m+1} \rVert^2_{L^2(\spacedomain \times \Omega)}$ across blocks. We consider three choices. A uniform partition simply sets $\tau^2_{b, uni} = \tau^2$, which targets the same relative error in every block:
\begin{equation}
    \tau^2 \lVert I_R^{m+1} \rVert^2_{L^2(\spacedomain \times \Omega)} = \sum_{b} \tau^2_{b,uni} \lVert I_{b,R}^{m+1} \rVert^2_{L^2(\spacedomain_b \times \Omega)},
\end{equation}
an intensity-weighted partition so that blocks with smaller intensities are allowed larger errors
\begin{equation}
    \tau^2 \lVert I_R^{m+1} \rVert^2_{L^2(\spacedomain \times \Omega)} = \sum_b \left( \tau^2 \frac{ \lVert I_R^{m+1} \rVert^2_{L^2(\spacedomain \times \Omega)} }{N \lVert I_{b,R}^{m+1} \rVert^2_{L^2(\spacedomain_b \times \Omega)}} \right) \lVert I_{b,R}^{m+1} \rVert^2_{L^2(\spacedomain_b \times \Omega)} = \sum_b \tau^2_{b,int} \lVert I_{b,R}^{m+1} \rVert^2_{L^2(\spacedomain_b \times \Omega)},
\end{equation}
a volume and intensity weighted partitioning so that blocks with small volume require tighter tolerances 
\begin{equation}
    \tau^2 \lVert I_R^{m+1} \rVert^2_{L^2(\spacedomain \times \Omega)} = \sum_b \left( \tau^2 \frac{\Delta V_b}{|\spacedomain|}
    \frac{\lVert I_R^{m+1} \rVert^2_{L^2(\spacedomain \times \Omega)} }{\lVert I_{b,R}^{m+1} \rVert^2_{L^2(\spacedomain_b \times \Omega)}} 
    \right)
    \lVert I_{b,R}^{m+1} \rVert^2_{L^2(\spacedomain_b \times \Omega)} = \sum_b \tau^2_{b,vol} \lVert I_{b,R}^{m+1} \rVert^2_{L^2(\spacedomain_b \times \Omega)} ,
\end{equation}

Finally, $\lVert I_{b,R}^{m+1}\rVert_{L^2(\spacedomain_b \times \Omega)}$ and the corresponding gradient norm are not computable in practice. We therefore replace them by quantities in which only the finite-volume error is incurred during the current step. With this substitution, the problem becomes to choose $h_b^2$ and $\epsilon_b^2$ so that
\begin{equation}
   \left( 2 C_{FV}^2 h_b^2 \lVert \lvert \nabla \hat{I}_{b,R}^m \rvert\rVert^2_{L^2(\spacedomain_b \times\Omega)} 
   +  \epsilon_b^2 \lVert \tilde{I}_{b,R}^{m+1}  \rVert^2_{L^2(\spacedomain_b \times \Omega)}
   \right) \leq \frac{\tau_{b,a}^2}{2} \lVert \tilde{I}^{m+1}_{b,R} \rVert^2_{L^2(\spacedomain_b \times \Omega)} \quad a \in \{uni, int, vol\}
\end{equation}
To satisfy this error tolerance we require the rounding tolerance to satisfy
\begin{equation}
    \epsilon_b^2 \leq  \frac{1}{2}  \tau_{b,a}^2  - 2 C_{FV}^2 h_b^2 \frac{\lVert \lvert \nabla \hat{I}_{b,R}^m \rvert\rVert^2_{L^2(\spacedomain_b \times\Omega)}}{  \lVert \tilde{I}^{m+1}_{b,R} \rVert^2_{L^2(\spacedomain_b \times \Omega)}}. \label{eq:rounding_bound}
\end{equation}
Thus, the rounding tolerance must be smaller than the local tolerance $\tau_{b,a}$ to account for the additional error incurred by the finite-volume scheme. If that contribution were zero, the rounding tolerance would be of the same order as the desired tolerance, as expected. Table~\ref{tab:rounding_tolerances} records the resulting expressions for the three partitioning strategies in terms of computable discrete quantities. In particular, uncomputable terms such as $\nabla I^m$ are replaced by computable surrogates such as $\nabla \hat{I}^m_{b,R}$.
\begin{table} 
\centering
\caption{Rounding tolerance upper bound for three different relative error partitions across the blocks, where $\tau \in \{\tau_{uni}, \tau_{int}, \tau_{vol}\}$.}
\label{tab:rounding_tolerances}
\begin{tabular}{lc}
\hline
Partition type & Rounding upper bound \\
\hline
\hline
Uniform &   $\frac{1}{2}\tau^2 - 2 C_{FV}^2 h_b^2 \frac{\lVert \lvert  \nabla \hat{I}_{b,R}^m \rvert \rVert^2_{L^2(\spacedomain_b \times \Omega)} } {\lVert \tilde{I}^{m+1}_{b,R} \rVert^2_{L^2(\spacedomain_b \times \Omega)}}$ \\
Intensity &  
$\frac{\tau^2}{2N} \frac{\sum_{b^\prime} \Delta V_{b^\prime}\lVert \tilde{I}_{b^\prime,R}^{m+1} \rVert^2_{L^2(\spacedomain_{b^\prime} \times \Omega)}}{\Delta V_b\lVert \tilde{I}^{m+1}_{b,R}\rVert^2_{L^2(\spacedomain_b \times \Omega)}} 
- 2 C_{FV}^2 h_b^2 \frac{\lVert \lvert \nabla \hat {I}_{b,R}^m \rvert \rVert^2_{L^2(\spacedomain_b \times \Omega)}}{\lVert \tilde{I}^{m+1}_{b,R}\rVert^2_{L^2(\spacedomain_b \times \Omega)}}$ \\
Volume-Intensity &  
$ \frac{\tau^2}{2|\spacedomain|} \frac{\sum_{b^\prime} \Delta V_{b^\prime}\lVert \tilde{I}_{b^\prime,R}^{m+1} \rVert^2_{L^2(\spacedomain_{b^\prime} \times \Omega)}}{\lVert \tilde{I}_{b,R}^{m+1} \rVert^2_{L^2(\spacedomain_b \times \Omega)}} 
- 2 C_{FV}^2 h_b^2 \frac{\lVert \lvert \nabla \hat {I}_{b,R}^m \rvert \rVert^2_{L^2(\spacedomain_b \times \Omega)}}{\lVert \tilde{I}_{b,R}^{m+1} \rVert^2_{L^2(\spacedomain_b \times \Omega)}} $ \\
\hline
\end{tabular}
\end{table}

After rounding for a given step, we can relate the refinement criteria to the chosen error tolerances. Let $\alpha \in (0,1)$ denote the fraction of the error budget assigned to the finite-volume scheme. We then seek to enforce
\begin{align}
    2C_{FV}^2 h_b^2 \lVert \lvert \nabla \hat{I}_{b,R}^m \rvert\rVert^2_{L^2(\spacedomain_b \times\Omega)}  &\leq \alpha \frac{\tau_{b,a}^2}{2} \lVert \tilde{I}^{m+1}_{b,R} \rVert^2_{L^2(\spacedomain_b \times \Omega)} \qquad \textrm{ and }  \label{eq:refinement_criteria} \\  
    \epsilon_b^2 \lVert \tilde{I}_{b,R}^{m+1}  \rVert^2_{L^2(\spacedomain_b \times \Omega)} &\leq (1-\alpha)\frac{\tau_{b,a}^2}{2} \lVert \tilde{I}^{m+1}_{b,R} \rVert^2_{L^2(\spacedomain_b \times \Omega)}, 
\end{align}
for  $a \in \{uni, int, vol\}.$ Equation~\eqref{eq:refinement_criteria} therefore suggests refinement when
\begin{equation}
     C_{FV}^2 h_b^2 \frac{\lVert \lvert \nabla \hat{I}_{b,R}^m \rvert\rVert^2_{L^2(\spacedomain_b \times\Omega)}}{\lVert \tilde{I}^{m+1}_{b,R} \rVert^2_{L^2(\spacedomain_b \times \Omega)}} > \alpha \frac{\tau_{b,a}^2}{4}.
\end{equation}
If this criterion is satisfied, the rounding tolerance can be restricted to
\begin{equation}
    \epsilon_b^2 \leq  \frac{1-\alpha}{2}  \tau_{b,a}^2. 
\end{equation}
The resulting simplified rounding tolerance bounds are provided in Table~\ref{tab:rounding_tolerances2}.
\begin{table} 
\centering
\caption{Rounding tolerance upper bound assuming spatial refinement enforces an $\alpha$-fraction of the desired tolerance, where $\tau \in \{\tau_{uni}, \tau_{int}, \tau_{vol}\}$.}
\label{tab:rounding_tolerances2}
\begin{tabular}{lc}
\hline
Partition type & Rounding upper bound \\
\hline
\hline
Uniform &   $\frac{1-\alpha}{2}\tau^2$ \\
Intensity &  
$\frac{(1-\alpha)\tau^2}{2} \frac{\sum_{b^\prime} \Delta V_{b^\prime}\lVert \tilde{I}_{b^\prime,R}^{m+1} \rVert^2_{L^2(\spacedomain_{b^\prime} \times \Omega)}} { N \Delta V_b\lVert \tilde{I}^{m+1}_{b,R}\rVert^2_{L^2(\spacedomain_b \times \Omega)}} $ \\
Volume-Intensity &  
$ \frac{(1-\alpha)\tau^2}{2} \frac{\sum_{b^\prime} \Delta V_{b^\prime}\lVert \tilde{I}_{b^\prime,R}^{m+1} \rVert^2_{L^2(\spacedomain_{b^\prime} \times \Omega)}} { |\spacedomain| \lVert \tilde{I}_{b,R}^{m+1} \rVert^2_{L^2(\spacedomain_b \times \Omega)}} $ \\ 
\hline
\end{tabular}
\end{table}
The same bound can be used to define a coarsening criterion. Let $\beta \in (0,\alpha)$, and coarsen when the block is sufficiently accurate:
\begin{equation}
     C_{FV}^2 h_b^2 \frac{\lVert \lvert \nabla \hat{I}_{b,R}^m \rvert\rVert^2_{L^2(\spacedomain_b \times\Omega)}}{\lVert \tilde{I}^{m+1}_{b,R} \rVert^2_{L^2(\spacedomain_b \times \Omega)}} < \beta\frac{\tau_{b,a}^2}{4}.
\end{equation}

\subsubsection{Interlinked criteria}

The interlinked adaptive strategy distributes a prescribed per-timestep error budget between spatial discretization error and rounding error. A fraction $\alpha$ of the local budget is assigned to the finite-volume contribution, while the remaining fraction $1-\alpha$ is assigned to rounding. The spatial component is monitored through the indicator \[
\eta_b
=
C_{FV}^2\, h_b^2\,
\frac{\left\|\nabla \hat{I}_{b,R}^m\right\|_{L^2(\Omega_b\times\mathbb{S}^2)}^2}
{\left\|\tilde{I}_{b,R}^{m+1}\right\|_{L^2(\Omega_b\times\mathbb{S}^2)}^2},
\] which serves as a proxy for the local finite-volume error and depends on the block mesh size and the local solution gradient. If $\eta_b$ exceeds a refinement threshold, the block is refined; if it falls below a coarsening threshold, the block becomes eligible for coarsening. In this way, the interlinked mechanism couples mesh adaptation directly to the local balance between spatial and rounding errors.

To assess this mechanism, we consider the 1D hohlraum test \citep[see, e.g.,][]{mocmc}.  In this problem, an infinite planar source emits isotropically into vacuum, yielding a time-dependent analytic solution determined by which propagation directions have reached each position at a given time. We invoke a 1D spatial domain $x\in[0,1]$, discretized initially with $N_x=40$ cells. The angular discretization is taken to be sufficiently fine, with $N_\theta=512$ and $N_\phi=1024$, so that angular discretization errors do not dominate the error budget. The interlinked refinement strategy is used with $C_{FV}=0.1$, and we examine two choices of the parameters $(\alpha,\beta)$, where $\alpha$ controls the fraction of the error budget allocated to the finite-volume contribution and $\beta$ sets the coarsening tolerance.

\begin{figure}
    \centering
    \includegraphics[width=1.0\linewidth]{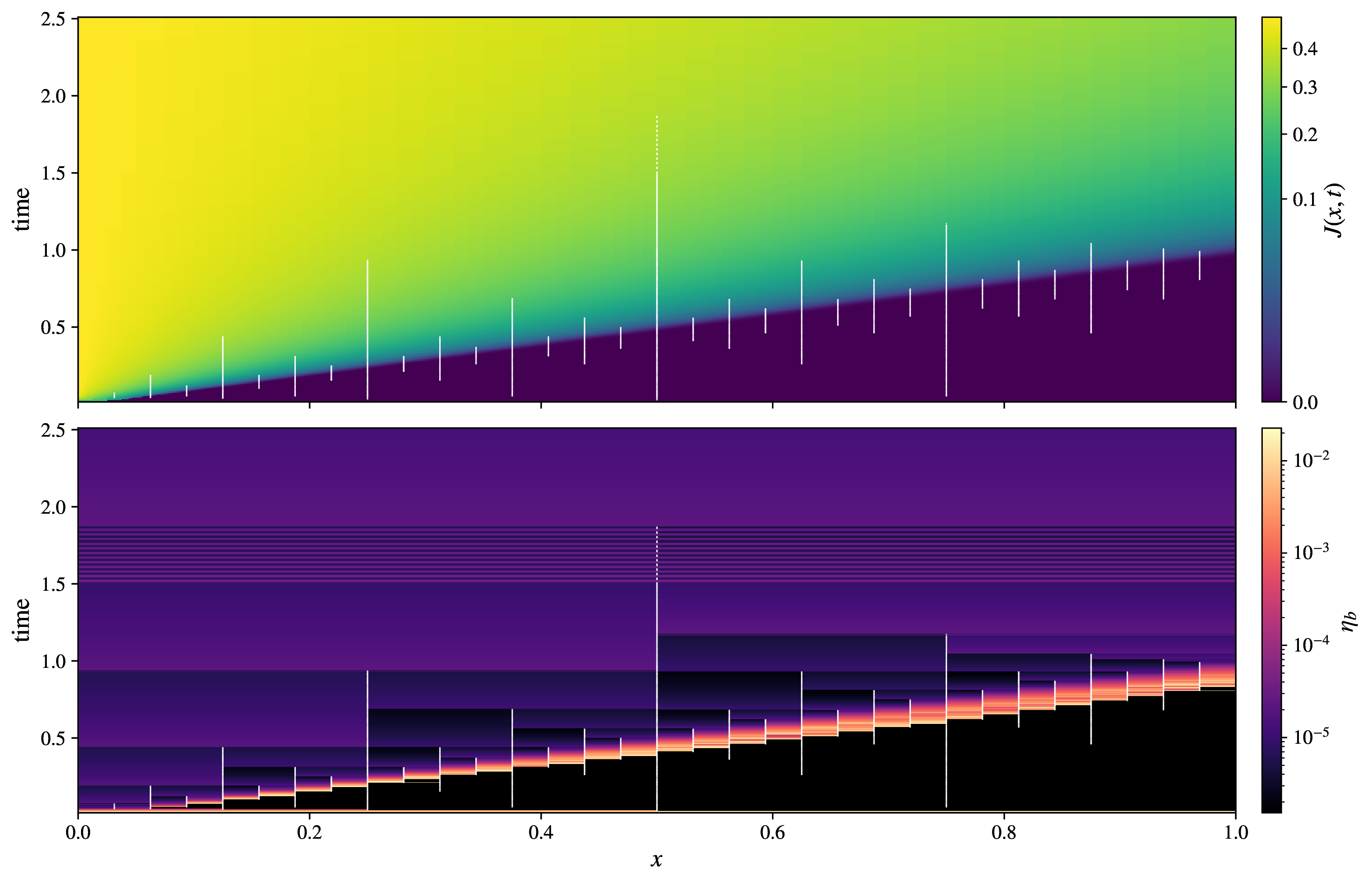}
    \caption{Interlinked refinement criterion for the 1D hohlraum problem with $(\alpha,\beta)=(0.9,0.4)$. The top panel shows the mean intensity, and the bottom panel the indicator $\eta_b$. Vertical white lines denote block boundaries; the appearance and disappearance of lines correspond to refinement and coarsening, respectively. Blocks near the advancing radiation front are refined, while those behind the front are coarsened. Refinement occurs when $\eta_b>\frac{\alpha}{4}\tau_b^2=2.25\times10^{-5}$, and coarsening when $\eta_b<\frac{\beta}{4}\tau_b^2=1.0\times10^{-5}$ with $\tau_b=10^{-2}$.}
    \label{fig:eta_plot}
\end{figure}

Figure \ref{fig:eta_plot} illustrates the behavior of the interlinked criterion for the 1D hohlraum problem with $(\alpha,\beta)=(0.9,0.4)$. The top panel shows the mean intensity in space and time, while the bottom panel shows the indicator $\eta_b$. The vertical white lines denote block boundaries; the appearance of new lines corresponds to refinement, while their disappearance corresponds to coarsening. The solution plot shows that blocks near the advancing solution front are refined, whereas the mesh coarsens after the front has passed. In this case, refinement is triggered when
\[
\eta_b > \frac{\alpha}{4}\tau_b^2 = 2.25\times10^{-5},
\]
while coarsening is permitted when
\[
\eta_b < \frac{\beta}{4}\tau_b^2 = 1.0\times10^{-5},
\]
using $\tau_b^2=(10^{-2})^2=10^{-4}$. The $\eta_b$ plot demonstrates the adaptive refinement criterion. Blocks whose indicator exceeds the refinement threshold are subdivided into two child blocks. This is evident during the first two time steps, where horizontal bright bands of large $\eta_b$ are followed by vertical white lines in the subsequent time step, marking the refined block boundaries. Conversely, when $\eta_b$ falls below the coarsening threshold, a refined block may merge with its siblings, provided that all children in the sibling family satisfy the coarsening criterion.

\begin{figure}
    \centering
    \includegraphics[width=1.0\linewidth]{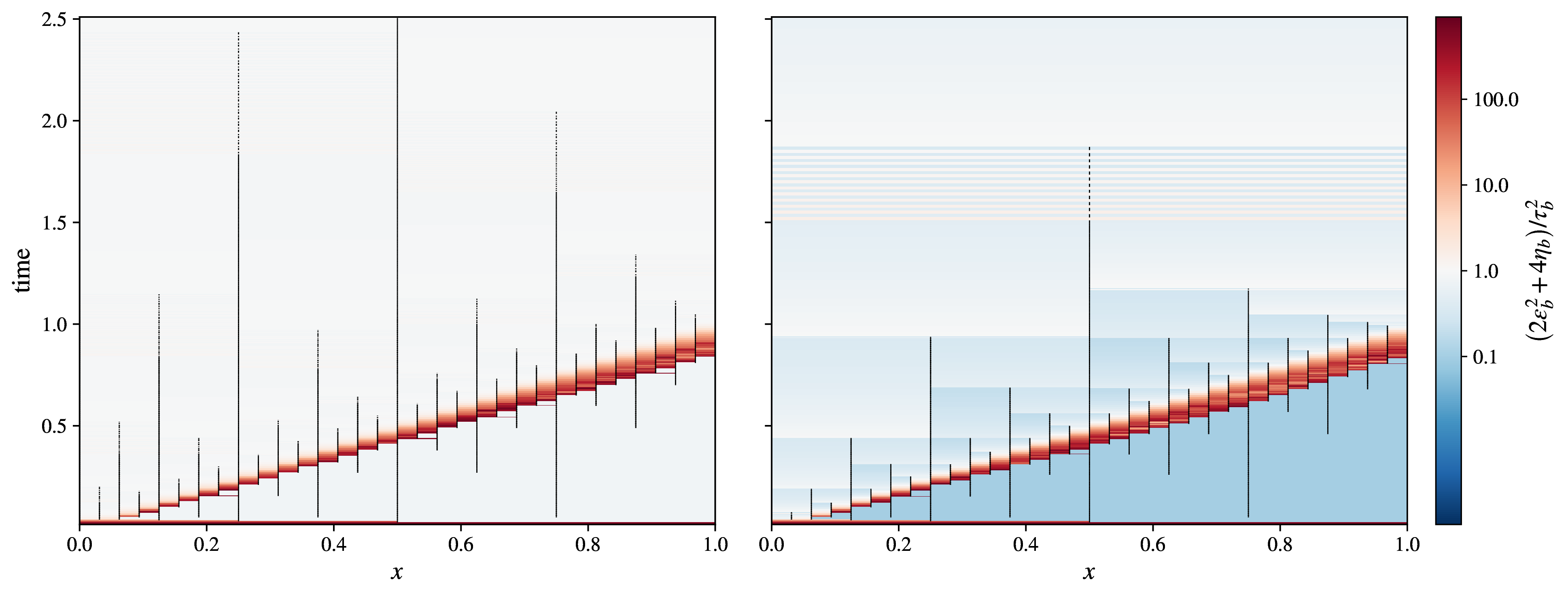}
    \caption{Normalized total error contribution, $(2\epsilon_b^2+4\eta_b)/\tau_b^2$, for $(\alpha,\beta)$ $=(0.2, 0.1)$ (left) and $(0.9, 0.4)$ (right); black vertical lines denote block boundaries. The quantity measures the combined rounding and finite-volume errors relative to the local error budget and is ideally below unity. Large values at early times arise from the advancing radiation front, while adaptive refinement reduces the spatial contribution as the solution evolves. In both cases, the ratio eventually falls below one. For larger $\alpha$, a greater share of the error budget is allocated to the spatial discretization, causing the normalized contribution to fall well below one at late times as $\eta_b$ decreases toward steady state.}
    \label{fig:total_error_comparison}
\end{figure}

Figure \ref{fig:total_error_comparison} shows the normalized quantity $(2\epsilon_b^2+4\eta_b)/\tau_b^2$
for $(\alpha,\beta)$ $=(0.2, 0.1)$ and $(0.9, 0.4)$ with the black vertical lines denoting the blocks this time. This quantity measures the combined rounding and finite-volume contributions relative to the local squared error budget $\tau_b^2$, and ideally it should remain below unity. At early times, the ratio is large because of the sharp gradients generated as radiation enters from the left boundary. Since the rounding tolerance is fixed,
the adaptive algorithm can reduce the total contribution only by controlling the spatial component through refinement and coarsening. In both parameter settings, the normalized total contribution eventually drops below one. For larger $\alpha$, the ratio falls well below one at late times because a greater fraction of the error budget is initially allocated to the spatial component. As the solution approaches a near-steady state, $\eta_b$ decreases substantially, leaving much of the spatial error budget unused.

A point worth emphasizing is that, although refinement is intended to reduce the finite-volume error, the refinement depth is capped by a user-prescribed maximum level. In the present example, even after a block reaches the maximum refinement level, the indicator $\eta_b$ may still remain above the refinement threshold. In such cases, further reduction of $\eta_b$ cannot be achieved through mesh refinement alone, and coarsening is only observed later, once the underlying solution dynamics naturally decrease the indicator.

Another important feature of the current implementation is that the rounding tolerance assigned to each block remains fixed throughout the simulation as a prescribed fraction, $1-\alpha$, of the total local error budget. By contrast, the finite-volume contribution $\eta_b$ is transient, and refinement/coarsening is used to keep it near its target fraction, $\alpha$, of the budget. However, in the above problem, at later times $\eta_b$ becomes significantly smaller than its allocated share, at which point coarsening is triggered in an attempt to raise the spatial error contribution, as observed in this problem. Once a block has reached the maximum admissible coarsening level, $\eta_b$ can no longer be increased, and the method consequently uses substantially less of the total budget than originally allocated. Although we do not pursue this here, the unused portion of the spatial budget could in principle be reassigned to rounding, potentially yielding additional compression.

\section*{Acknowledgements}

A.A.G and A.D were supported by the Los Alamos National Laboratory (LANL) under the project ``Algorithm/Software/Hardware Co-design for High Energy Density applications'' at
the University of Michigan. P.D.M, J.C.D, C.D.M., J.M.M., and L.F.R. were supported by LANL under the M$^3$AP project.  We used Claude to assist with language editing, organization, and clarity of selected manuscript sections. The authors reviewed and revised all AI-assisted text and are fully responsible for the accuracy, originality, citations, analyses, and conclusions presented in this work.  This research used resources provided by the Darwin testbed at LANL which is funded by the Computational Systems and Software Environments subprogram of LANL's Advanced Simulation and Computing program (NNSA/DOE). LANL is operated by Triad National Security, LLC, for the National Nuclear Security Administration of U.S. Department of Energy (Contract No. 89233218CNA000001). This document is approved for unlimited release under LA-UR-26-24582.

\bibliographystyle{aasjournalv7}
\bibliography{refs}

\end{document}